\documentclass[a4paper,fleqn,usenatbib]{mnras}
\usepackage{newtxtext, newtxmath, graphicx,enumerate,amsmath, morefloats,amstext, amssymb,array,multirow}
\usepackage[T1]{fontenc}
\usepackage{ae}
\usepackage{aecompl}
\usepackage[figure,figure*]{hypcap}

\newcommand{\heiw}{\mbox{\ion{He}{i} $\lambda$5876}}

\newcommand{\neiiiw}{\mbox{[\ion{Ne}{iii}] $\lambda$3869}}
\newcommand{\neiiibw}{\mbox{[\ion{Ne}{iii}] $\lambda$3967}}

\newcommand{\oiw}{\mbox{[\ion{O}{i}] $\lambda$6300}}
\newcommand{\oiiw}{\mbox{[\ion{O}{ii}] $\lambda$3727}}

\newcommand{\niibw}{\mbox{[\ion{N}{ii}] $\lambda$6583}}
\newcommand{\niiw}{\mbox{[\ion{N}{ii}] $\lambda \lambda$6548,6583}}
\newcommand{\niitw}{\mbox{[\ion{N}{ii}] $\lambda$5755}}
\newcommand{\siiaw}{\mbox{[\ion{S}{ii}] $\lambda$6716}}
\newcommand{\siibw}{\mbox{[\ion{S}{ii}] $\lambda$6731}}
\newcommand{\siiw}{\mbox{[\ion{S}{II}] $\lambda \lambda$6716,6731}}

\newcommand{\oiiibw}{\mbox{[\ion{O}{iii}] $\lambda$5007}}
\newcommand{\oiiiw}{\mbox{[\ion{O}{III}] $\lambda \lambda$4959,5007}}
\newcommand{\oiiitw}{\mbox{[\ion{O}{III}] $\lambda$4363}}

\newcommand{\siitw}{\mbox{[\ion{S}{II}] $\lambda\lambda$4068,4076}}
\newcommand{\oiitw}{\mbox{[\ion{O}{II}] $\lambda\lambda$7320,7330}}
\newcommand{\feiiaw}{\mbox{[\ion{Fe}{II}] $\lambda$4287}}
\newcommand{\feiibw}{\mbox{[\ion{Fe}{II}] $\lambda$4359}}
\newcommand{\caiiaw}{\mbox{[\ion{Ca}{II}] $\lambda$7291}}
\newcommand{\caiibw}{\mbox{[\ion{Ca}{II}] $\lambda$7324}}
\newcommand{\naiw}{\mbox{\ion{Na}{I} $\lambda\lambda$5890,5896}}

\newcommand{\oi}{\mbox{[\ion{O}{i}]}}
\newcommand{\oii}{\mbox{[\ion{O}{ii}]}}
\newcommand{\nii}{\mbox{[\ion{N}{ii}]}}
\newcommand{\hal}{\mbox{H$\alpha$}}
\newcommand{\sii}{\mbox{[\ion{S}{ii}]}}
\newcommand{\hb}{\text{H$\beta$}}
\newcommand{\hg}{\text{H$\gamma$}}
\newcommand{\hd}{\text{H$\delta$}}
\newcommand{\heps}{\text{H$\epsilon$}}
\newcommand{\hz}{\text{H$\zeta$}}
\newcommand{\oiii}{\text{[\ion{O}{iii}]}}
\newcommand{\feii}{\mbox{[\ion{Fe}{ii}]}}

\title[Warm Ionized Gas Temperature in Quiescent Galaxies]{Shocks or Photoionization: Direct Temperature Measurements of the Low-Ionization Gas in Quiescent Galaxies}
\author[R. Yan]{Renbin Yan$^{1}$\thanks{E-mail: yanrenbin@uky.edu}
\\
$^{1}$Department of Physics and Astronomy, University of Kentucky, 505 Rose Street, Lexignton, KY, 40506; yanrenbin@uky.edu}

\date{Accepted XXX. Received YYY; in original form ZZZ}
\pubyear{2017}

\begin{document}
\label{firstpage}
\pagerange{\pageref{firstpage}--\pageref{lastpage}}
\maketitle

\begin{abstract}
The ionization mechanism of the low-ionization gas in quiescent red sequence galaxies has been a long-standing puzzle. Direct temperature measurements would put strong constraints on this issue. We carefully selected a sample of quiescent red sequence galaxies from SDSS. We bin them into three bins with different \nii/\hal\ and \nii/\oii\ ratios, and we measure the temperature-sensitive \oiiitw, \niitw, \siitw, and \oiitw\ lines in the stacked spectra. { The \sii\ doublet ratios indicate the line-emitting gas is in the low density regime ($\sim10-200$ cm$^{-3}$).} We found the temperatures in the S$^+$ zones to be around 8000K, the temperatures in the O$^+$ zone to be around $1.1-1.5\times10^4$K, and the temperatures in the N$^+$ zones to be around $1-1.4\times10^4$K. { The \oiiitw\ line is not robustly detected.}
We found that the extinction corrections derived from Balmer decrements would yield unphysical relationships between the temperatures of the S$^+$ zones and O$^+$ zones, indicating that the extinction is significantly overestimated by the measured Balmer decrements. We compared these line ratios with model predictions for three ionization mechanisms: photoionization by hot evolved stars, shocks, and turbulent mixing layers. { For the photoionization and shock models, the hot temperatures inferred from \sii\ and \nii\ coronal-to-strong line ratios require metallicities to be significantly subsolar. However, the \nii/\oii\ line ratios require them to be supersolar. None of the models could simultaneously explain all of the observed line ratios, neither could their combinations do.}
\end{abstract}




\section{Introduction} 

Contrary to the conventional wisdom, recent studies have shown that elliptical and lenticular galaxies also have interesting interstellar medium. They not only have X-ray-emitting hot gas, many of them also have substantial amount of warm and cold gas, and dust \citep[e.g.][]{Phillips86,Kim89,Buson93,Goudfrooij94,Macchetto96,Zeilinger96,Lauer05,Sarzi06,DavisT11, Singh13, Gomes16}.
The evidence for the warm gas in early-type galaxies come from optical emission line measurements. Unlike late-type galaxies, in most cases, the emission lines are not due to star formation. Most early-type galaxies with line emission show a line ratio pattern similar to Low-ionization Nuclear Emission-line Regions (LINER, \citealt{Heckman80}). This often leads to them being classified as a type of active galactic nuclei. However, many recent studies have shown that these line emission is not only spatially-extended, but have a spatial gradient in line ratio that are consistent with being photoionized by sources that are spatially distributed like the stars, rather than a central AGN \citep{Sarzi10, YanB12, Singh13, Belfiore15, Gomes16}. Therefore, the name LINER is no longer appropriate. In some recent papers, these phenomena have been referred to as LINER-like. \cite{Belfiore15} recently has renamed them as Low-Ionization Emission Regions or LIERs for short. LINER-like and LIER are both referring to the same phenomenon, namely the spatially-extended optical emission lines observed in a large fraction of quiescent red galaxies where the line ratios do not match those of star-forming regions. 
The exact ionization mechanism certainly could be different in individual galaxies. But for the majority of early-type galaxies where it is dominated by extended emission, having uniform line ratio and similar equivalent width (EW), there is likely a common mechanism shared by most of them. Although the AGN photoionization model has been ruled out for the majority of low-ionization emission-line galaxy, there are several mechanisms that are spatially distributed, such as photoionization by hot evolved stars, collisional ionization by shocks, and turbulent mixing layers.

Photoionization by hot evolved stars is a potential explanation for these warm ionized gas. It was originally proposed by \cite{Binette94}, and further developed by \cite{Stasinska08} and \cite{CidFernandes11}. Several observational evidences favor this explanation. The strongest evidences are two. One is the reasonably tight correlation between emission-line surface brightness and the stellar surface density shown by \cite{Sarzi10}. The other is that the ionization parameter gradient measured statistically by \cite{YanB12} matches the prediction of a distributed ionizing source following the stellar density profile. The latter work also found the luminosity-dependence of the ionization parameter gradient also match the theoretical prediction. These seem to be very strong evidence. 

However, there are also serious problems associated with this explanation. 
First, the ionization photon budget does not work out. Current stellar evolution theory suggests the number of ionizing photons from an evolved stellar population is roughly $10^{41}$ per second per solar mass, and it changes little with age after 1Gyr. In order to power all the line emission, the gas has to absorb 100\% of the ionizing photons from post-AGB stars. This would require the gas to surround each individual stars. However, in many galaxies, the gas are observed to have different kinematics from the stars \citep{Sarzi06, DavisT11, Gomes16}, suggesting an external origin. If the neutral gas originated externally, we do not expect them to provide complete covering fraction. Thus, we cannot explain the level of emission given the current prediction of the stellar evolution theory. Second, if the gas originated externally and is randomly positioned relative to the hot evolved stars, we will also have trouble to produce the kind of ionization parameter that are seen. For detailed calculations, see \cite{YanB12}. 

Shocks are another popular mechanism for the ionization. Given the amount of stellar mass loss and the large stellar velocity dispersion in these early-type galaxies, cloud-cloud collisions will produce shocks. But it is unclear whether it is the dominant emission-line luminosity contributor. In an early-type galaxy observed by the SDSS-IV MaNGA (Mappping Nearby Galaxies at Apache Point Observatory) survey \citep{Bundy15,Yan16b}, \cite{Cheung16} found narrow bi-symmetric features in its \hal\ equivalent width map. The proposed theory is that winds powered by a central radio AGN produces shocks in the direction of the winds, enhancing the line emission in those directions. 

Turbulent mixing layers may also produce the observed emission lines \citep{Slavin93}. Shear flows at the boundaries of hot and cold gas could produce gas at intermediate temperatures and give similar line ratios as observed in LIER and diffuse ionized gas in star-forming galaxies.  

One reliable way to distinguish these different ionization mechanisms is the temperature of the gas. Photoionized gas usually has a temperature around $10^4$K, while shocked gas and turbulent mixing has higher temperatures, approaching $10^5$K. If we can measure the temperature, we can tell which ionization mechanism is at work. There are a few temperature-sensitive line ratios in an optical spectrum, such as \oiiibw/\oiiitw, \niibw/\niitw. The weaker line in these line ratios are usually too weak to be detected in an SDSS-quality spectrum. In this paper, we measure these temperature-sensitive lines in carefully stacked spectra of quiescent red galaxies. This allows us to measure the gas temperature and put strong constraints on the ionization mechanisms. 

The paper is organized as the following. We describe the data in Section 2. The methods of sample selection, line measurements and zero-point corrections, selection of subsamples, stacking, stellar continuum subtraction, and weak line measurements in the stacked spectra are described in Section 3. We derive the extinction and gas temperatures in Section 4. We compare the results with models in Section 5 and conclude in Section 6. 

Throughout the paper, we assume a flat $\Lambda$CDM cosmology with $\Omega_m=0.3$ and a Hubble constant $H_0=100h$~km~s$^{-1}$ Mpc$^{-1}$. The magnitudes used are all in the AB system.

\section{Data}

Sloan Digital Sky Survey \citep{York00} has obtained five-band optical imaging in one quarter of the whole sky and obtained spectra for more than half a million galaxies in the local Universe. For this paper, we use the spectra from SDSS DR7 \citep{SDSSDR7}.  Our spectra are obtained from the Science Archive Server of SDSS and are read in by the code using the {\it readspec.pro} routine in the IDLSPEC2D package. We make use of New York University Value Added Catalog \citep{BlantonSS05} which is a magnitude-limited sample brighter than 17.9 in $r$-band. g

As we pointed out before in \cite{Yan11flux}, the standard flux calibration of the spectra show small residuals on large wavelength scales. For this work, we have applied the correction derived by \cite{Yan11flux} in each spectrum. We also correct the spectra for galactic extinction using the dust map measured by \cite{SchlegelFD98} and the extinction curve determined by \cite{O'Donnell94}.

\section{Methods}

\subsection{Selection of Quiescent Red Galaxies}

To measure the pure low-ionization emission line spectrum, we need a sample of quiescent red galaxies. By `quiescent', we are referring to galaxies that are not forming stars. Such low-ionization diffuse emission may also occur in star-forming galaxies, but it is overwhelmed by line emission from star-forming HII regions. In quiescent galaxies, we can study them uncontaminated.

\begin{figure}
\includegraphics[width=\columnwidth]{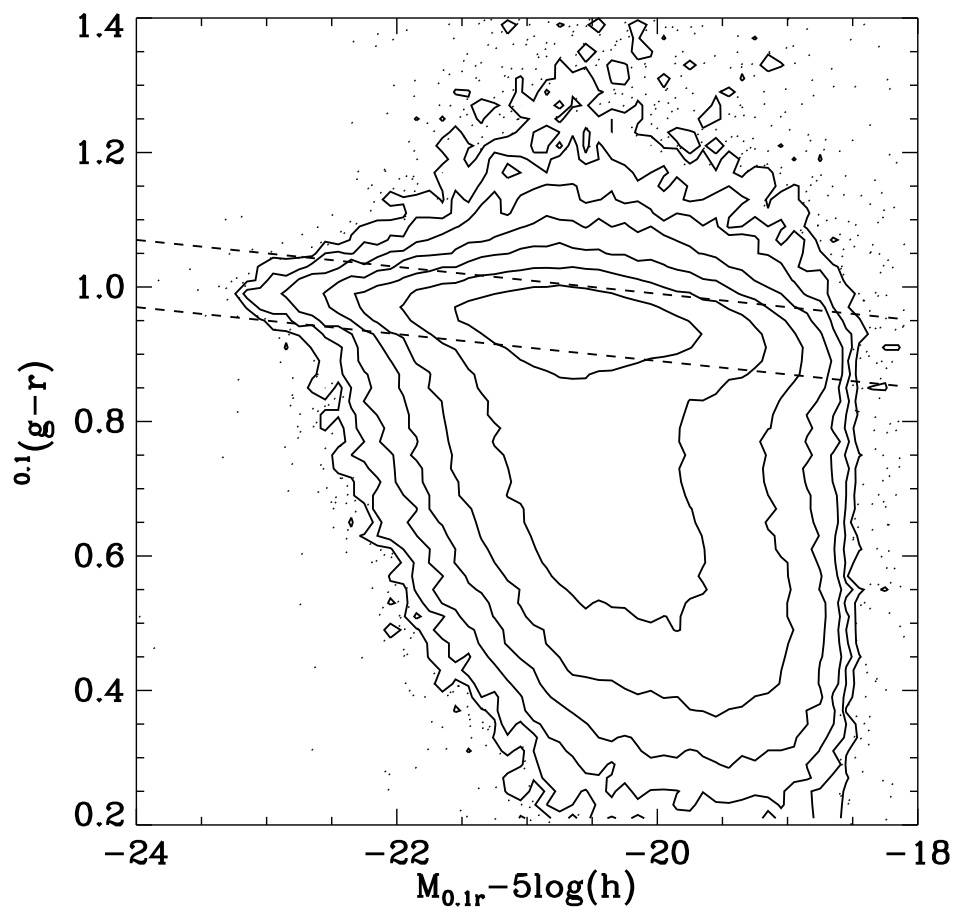}
\caption{colour-magnitude diagram for all SDSS main sample galaxies with $0.09<z<0.1$. The colour bi-modality can be clearly seen. The tilted lines indicate our thresholds for selecting red sequence galaxies.}
\label{fig:cmd}
\end{figure}

We choose a redshift range of $0.06<z<0.15$. We choose $z>0.06$ so that the 3\arcsec\ SDSS fibers cover a scale more than 2.5 kpc/h. As shown by \cite{YanB12}, on these scales, the total line emission luminosity is dominated by large scale emission, rather than the nuclear emission which may have contribution from AGN photoionization. We choose $z<0.15$ to ensure sufficient S/N for the majority of the sample.

First, we select a red galaxy sample from SDSS using a cut in colour-magnitude diagram.  Figure~\ref{fig:cmd} shows the $^{0.1}(g-r)$ vs. $M_{0.1r}-5\log h$ plot for all SDSS main sample galaxies in a narrow redshift bin. The colour bi-modality is evident in this figure. We apply empirical cuts to select the red galaxies in-between the two lines. The cuts are defined by the following inequalities.
\begin{eqnarray}
^{0.1}(g-r) &>& -0.02 (M_{^{0.1}r}-5\log h) +0.49 \\
^{0.1}(g-r) &<& -0.02 (M_{^{0.1}r}- 5\log h) +0.59
\end{eqnarray}

\begin{figure}
\begin{center}
\includegraphics[width=0.45\textwidth]{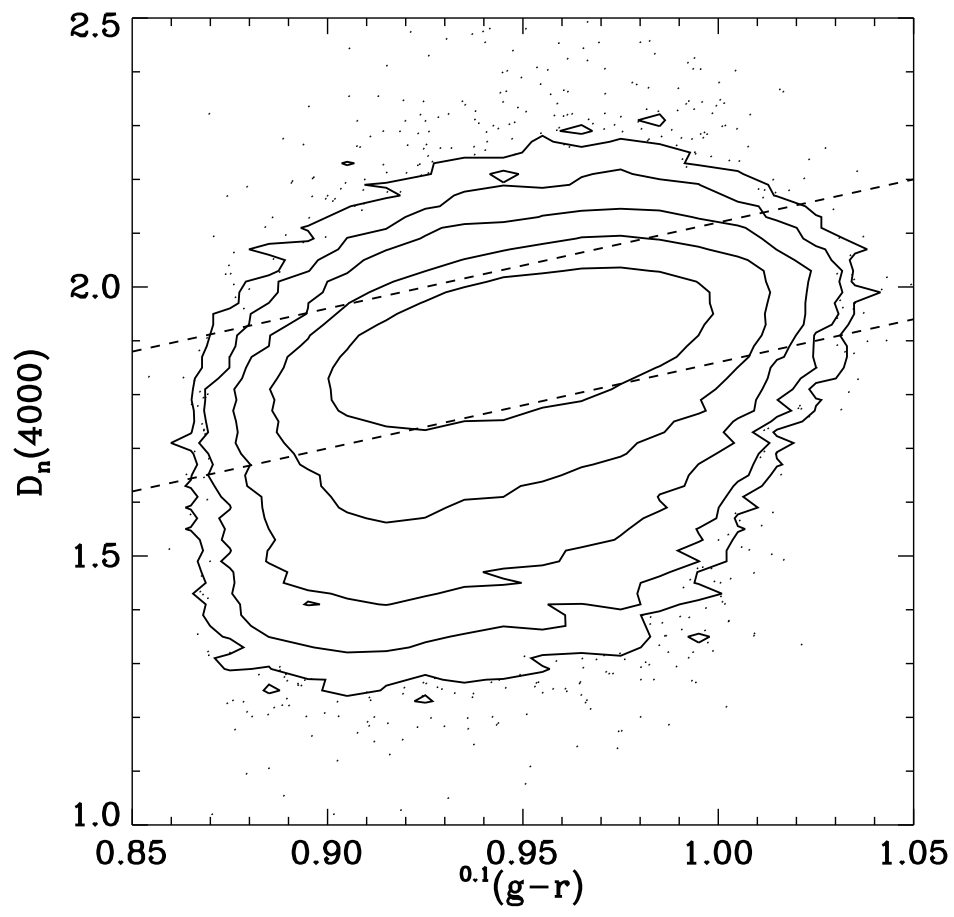}
\caption{D4000 vs. $^{0.1}g-r$ for red sequence galaxies at $0.09<z<0.1$. The tilted dashed lines indicate our cut for selecting the quiescent red sequence galaxies.}
\label{fig:d4000_gr}
\end{center}
\end{figure}

Red sequence galaxies can also include star-forming galaxies that appear red due to dust extinction. Here, we demonstrate that an additional cut based on $D_n(4000)$ is very effective in removing these dusty star-forming galaxies. Figure~\ref{fig:d4000_gr} shows the distribution of red sequence galaxies selected above in the $D_n(4000)$ vs. $g-r$ space. Our definition of $D_n(4000)$ is the same as those used in \cite{Balogh99} and \cite{KauffmannHT03}, and is defined as the ratio between the average flux density ($f_\nu$) between the two windows bracketing the $4000\AA$: $4000-4100\AA$ and $3850-3950\AA$. Figure~\ref{fig:d4000_gr} clearly shows a correlation between D4000 and g-r colour for the majority of red galaxies. But there are also galaxies with lower D4000 that fall below the correlation. 

\begin{figure}
\begin{center}
\includegraphics[width=0.45\textwidth]{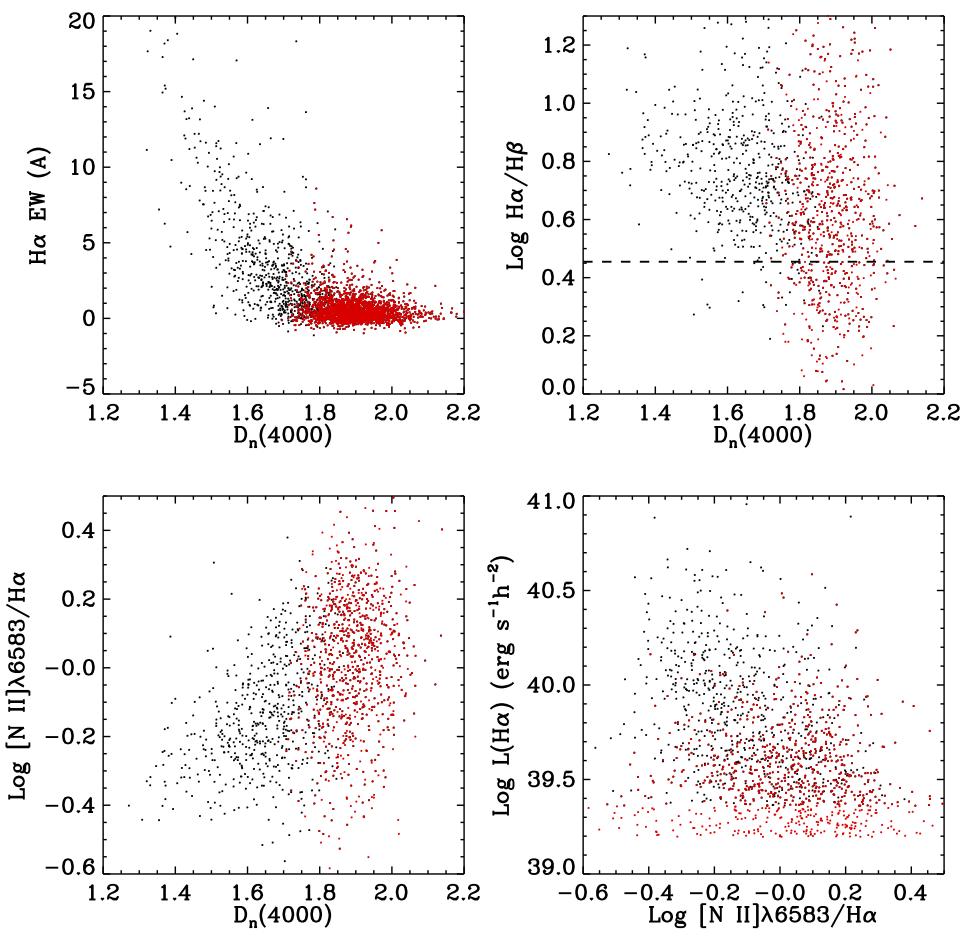}
\caption{Panel (a): $H\alpha$ emission equivalent width (EW) vs. $D_n(4000)$ for red sequence galaxies with $0.09<z<0.1$. Low $D_n(4000)$ galaxies have stronger emission.  (b): Balmer decrement for stronger line-emitting red-sequence galaxies in the same redshift range. { The dashed line indicates the Case B ratio for 10,000K.} Only the brightest 40\% of galaxies in \hal\ luminosity are plotted. Low $D_n(4000)$ red galaxies have higher dust extinction. (c) \nii/\hal\ ratio for red galaxies as a function of D4000. Low $D_n(4000)$ galaxies have lower \nii/\hal. (d) The \hal\ luminosity vs. \nii/\hal\ ratio distribution of low $D_n(4000)$ galaxies follow the expectation of a composite of low-ionization emission and star-forming region emission. In all panels, the red points indicate those galaxies above the lower cut shown in Figure~\ref{fig:d4000_gr}.}
\label{fig:dustyredgal}
\end{center}
\end{figure}

What are these galaxies? As shown in Fig.~\ref{fig:dustyredgal}, among all red galaxies, low $D_n(4000)$ ones have stronger line emission and a higher Balmer decrement than high $D_n(4000)$ ones, indicating that they are dusty star-forming galaxies. The younger the stellar population (lower $D_n(4000)$), the higher an extinction is needed to have a colour as red as an unextincted old population. The low $D_n(4000)$ galaxies also have lower \nii/\hal\ ratios, but not as low as pure star-forming galaxy. This intermediate line ratio is probably due to a combination of low-ionization emission and star formation, as pointed out by \cite{YanB12} in their Figure 2 and shown here again in panel (d) of Figure~\ref{fig:dustyredgal}.
Because the two windows defining $D_n(4000)$ are separated by only 150\AA, it is much less sensitive to dust extinction than $^{0.1}(g-r)$ colour, in which the two bands are separated by $\sim1300$\AA. Therefore, $D_n(4000)$ is better at reflecting the intrinsic property of the stellar population. 

Therefore, to exclude dusty star-forming galaxies from our analysis, we apply two cuts on $D_n(4000)$ { as a function of $^{0.1}(g-r)$ and select only galaxies in-between the two cuts}. The cuts we adopt are shown in Figure~\ref{fig:d4000_gr} and can be expressed by the following inequalities. 

\begin{eqnarray}
D_n(4000) &>& 1.6~^{0.1}(g-r) +0.26 \\
D_n(4000) &<& 1.6~^{0.1}(g-r) +0.52
\end{eqnarray}





\subsection{Emission line measurements in individual spectra}

We measure the strong emission lines(\oiiw, \hb\, \oiiibw, \oi, \hal, \niibw, \siiw) after subtracting the stellar continuum in each spectrum. We fit for the stellar continuum using a linear combination of templates. The templates are made with \cite{BC03} stellar population synthesis models with solar metallicity and span a range of ages. The fitting is done in the same way as described by \cite{Yan06}, except for the use of multiple templates. After the continuum subtraction, the residuals around the line may still have systematic difference from zero. We thus fit the residual continuum by a linear function in two sidebands around the line and then subtract off the linear fit from the residual spectrum. Finally, we sum the flux within narrow windows around each line to get the line flux. Note we do not use Gaussian-fitted flux because the Gaussian fitting is ill-behaved when the line flux is zero or very low. The center and sidebands windows for each emission line are listed in Table.~\ref{tab:linedef}. 
The \hb\ line flux is measured within a fairly narrow window out of concern of imperfect stellar continuum subtraction around \hb. As the narrow window does not always cover the complete line profile, we correct for the missing flux assuming a Gaussian profile with a dispersion equal to that measured for \hal. 

\begin{table*}
\begin{tabular}{llll}
\hline\hline
Line & Center window (\AA) & Left sideband(\AA) & Right sideband (\AA)\\ \hline
\oiiw & 3717.36---3739.36 & 3697.36---3717.36 & 3739.36---3759.36\\
\neiiiw & 3859.85---3879.85 & 3839.85---3859.85 & 3879.85---3899.85\\
\hb & 4857.46---4867.04 & 4798.88---4838.88 & 4885.62---4925.62\\
\oiiibw & 4998.2---5018.2 & 4978.2---4998.2 & 5018.2---5038.2 \\
\oiw & 6292.05---6312.05 & 6272.05---6292.05 & 6312.05---6332.05 \\
\hal & 6554.61---6574.61 & 6483---6538 & 6598---6653 \\
\niibw &6575.28---6595.28 & 6483---6538 & 6598---6653 \\
\siiw & 6705.48---6745.48 & 6687.48---6705.48 & 6745.48---6763.48 \\
\hline\hline
\end{tabular}
\caption{Definition of windows for strong emission line measurements in both individual and stacked spectra.}
\label{tab:linedef}
\end{table*}

\subsection{Zero-point correction for emission lines}

The galaxies we selected have very uniform continuum shapes and line ratios. The standard deviation between the \oii\ continuum level and the \hal\ continuum level is only 0.046 dex (10.6\%). 

Because the emission lines of these galaxies are relatively weak, it is extremely important to remove as much systematics as possible from the emission line measurements. Otherwise, the line ratios will be biased and there could be artificial correlation between line ratio and line strength (e.g. between \hb/\hal\ ratio and \hal\ strength, or between \oiii/\oii\ ratio and \oii\ strength). We noticed that there are systematics associated with our emission line EW measurements, because even for galaxies with every emission line other than \oiii\ to be negative, the median \oiii\ emission line flux is still having a positive value.
Another example is, for a sample of galaxies with nearly zero \hal\ EW, the \hb\ EW has a median of 0.16 A, which is not physical. 
This means that there is a zero-point offset in our measurement of line emission EW. Based on this logic, we first determine the zero-point offsets of all the emission line EW. The systematics are likely resulting from inaccurate stellar continuum subtraction, thus, they should be removed in EW rather than in flux. 

The logic we adopt is, that the EW of different emission lines should be positively correlated with each other, as shown by the EW vs. EW plots in Figure~\ref{fig:ew_vs_ew}. The true value of an emission line ``A'' should be among the lowest among all galaxies if all the other emission lines are among the lowest. In addition, we assume there should be a population of quiescent galaxies where all line emission is consistent with zero. There are several strong lines in the spectra: \hal, \niibw, \siiw, \oiiibw, and \oiiw. To get the EW zero-point for each strong line, we select a sample with the lowest emission line EWs in all the other strong lines except the one to be considered. 

\begin{figure*}
\begin{center}
\includegraphics[width=\textwidth]{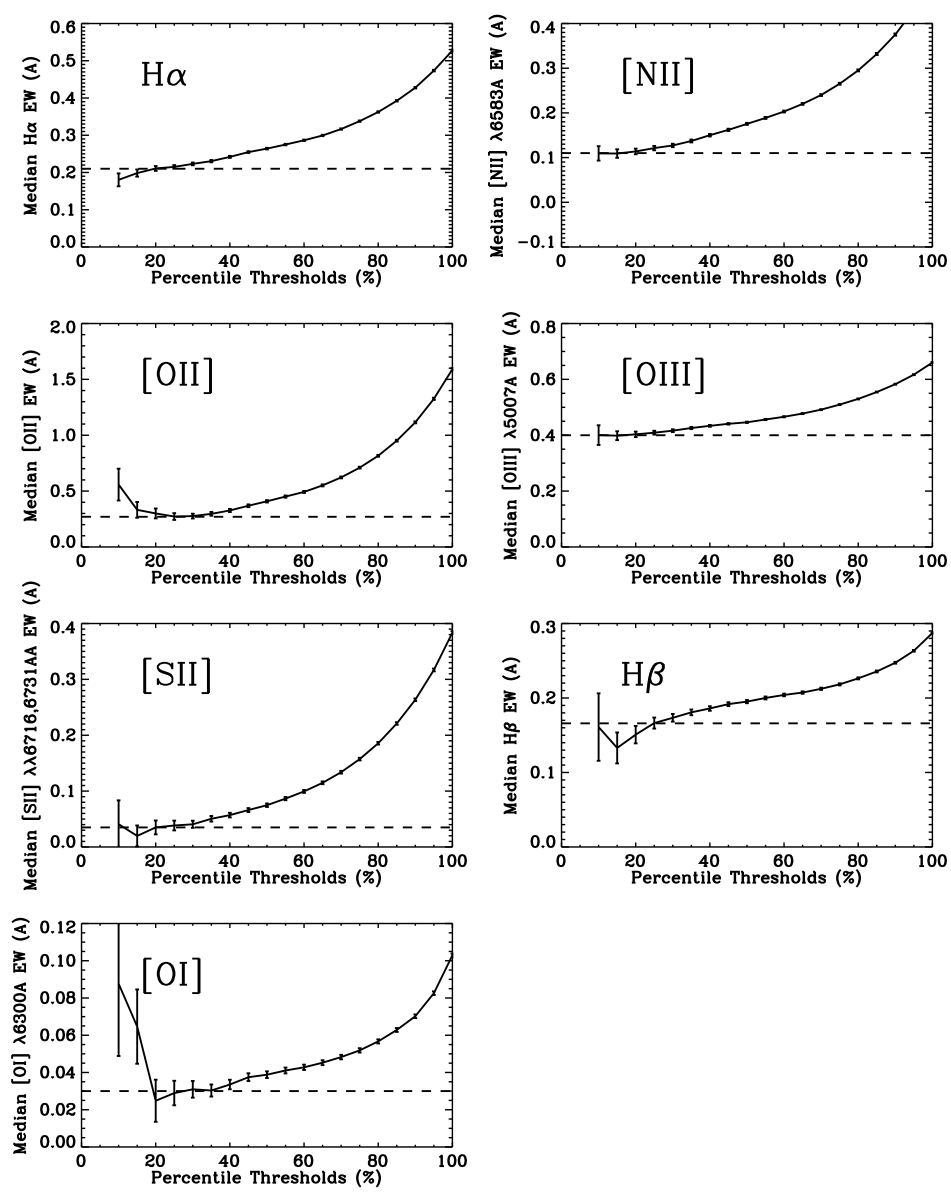}
\caption{Median EW for each emission line for subsamples with the lowest EWs in the other strong lines, plotted as a function of the percentile threshold for selecting such subsamples. Here, the EWs shown are before the zero-point correction is applied. The horizontal lines indicate the adopted zero-point correction for each line.}
\label{fig:zeropoint}
\end{center}
\end{figure*}

\begin{figure*}
\begin{center}
\includegraphics[width=\textwidth]{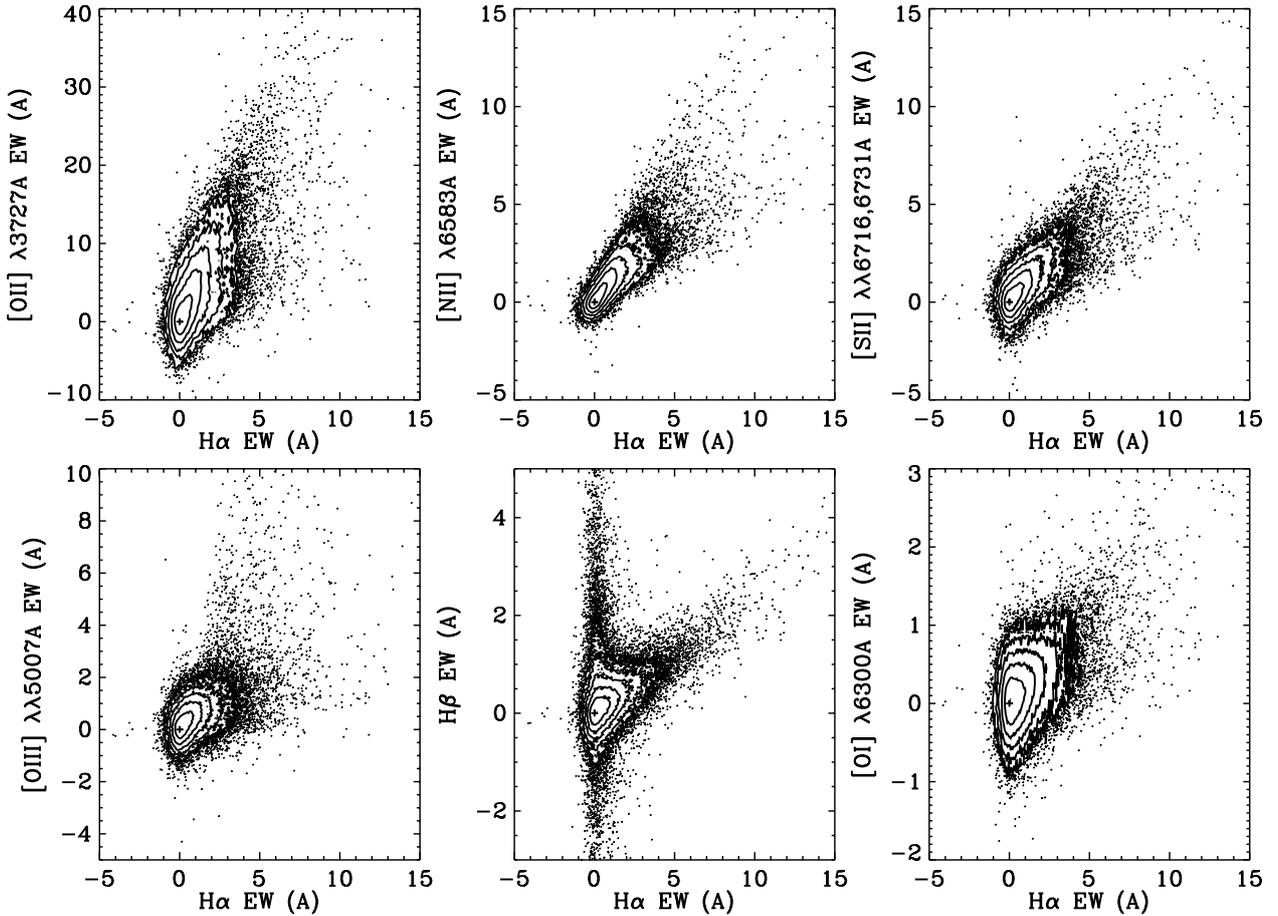}
\caption{Zero-point-corrected EWs of six different emission lines are compared to \hal\ EW for the high-D$_{\rm n}$(4000) red galaxy sample. All lines show positive correlation with \hal. The contour levels indicate the number density of points. They are equally-spaced in log density. The crosses indicate the location of the origin [0,0]. Before the zero-point correction, the highest density regions do not center around the origin.}
\label{fig:ew_vs_ew}
\end{center}
\end{figure*}

For example, to evaluate \oii\ EW zero-point, we first select all galaxies whose other strong lines (\hal, \nii, \oiii, and \sii) have EWs below the 10th percentile, respectively for each emission line, among all the quiescent galaxies. We measure the median \oii\ EW among these galaxies. We then increase the percentile threshold to 15th percentile, 20th percentile, etc. Finally, we plot the median \oii\ EW values as a function of the percentile thresholds in Fig.~\ref{fig:zeropoint}. As we go to higher percentile thresholds, the median \oii\ EW should monotonically increase. At low enough percentile thresholds, the median \oii\ EW should be consistent to being flat because those line measurements are completely due to noise. As shown in Figure~\ref{fig:zeropoint}, the median \oii\ EW first decreases and then increases. The initial decline is basically consistent with being flat. It declines because of noisy measurements and the small sample statistics. We thus set the zero-point of \oii\ EW to be the median value for the 25th percentile which is a good estimate for those galaxies whose line emission are consistent with being zero. 

We do this for all other strong lines. When we determine the zero-point for \hal\ or \nii, we exclude both of them in selecting the lowest EW samples (i.e., we only look at \oii, \oiii\ and \sii\ to select the subsample for each percentile threshold). Because \hal\ and \nii\ are measured using the same continuum subtraction, the two lines' EW zero-points could be correlated. If the line to be considered is relatively weak (\hb, \oi), then we look at all the strong lines to determine the sample for each percentile threshold. 

We took the values of the median EWs for the 25-th percentile subsamples as the zero-point for all the lines. These values are subtracted from the original EW measurements. Afterwards, we recompute the correct line fluxes by multiplying the zero-point-corrected EWs with their corresponding continuum levels in each spectrum. 

After this correction to the EWs, we plot the EWs of the lines against each other (Fig.~\ref{fig:ew_vs_ew}).  With the correction, there is a population of galaxies whose distribution in every emission line centers around the origin. These galaxies are consistent with having zero line emission. 

Note that this method assumes the intrinsic emission line EWs for those galaxies with the weakest lines are zero. However, it is entirely possible for all early type galaxies to have an intrinsic minimum line strength. We think this is less likely. Even if there were such a mechanism to give every galaxy a positive line strength, our method below would remove the effect of those constant emission EW and only probe the line emission above this constant minimum level. 

\subsection{Constructing the strong-line subsamples}
\begin{figure}
\begin{center}
\includegraphics[width=0.45\textwidth]{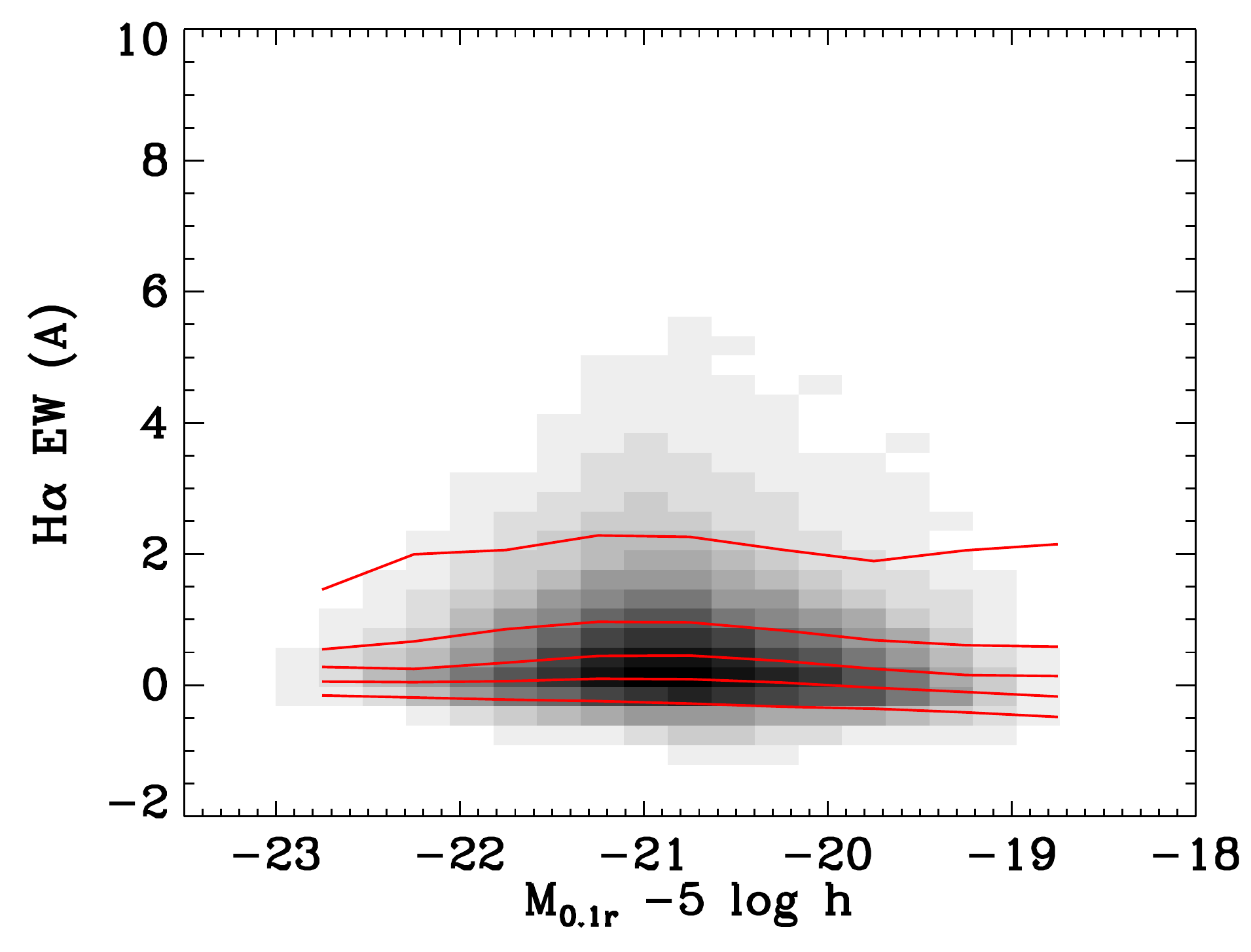}
\caption{This figure shows that the equivalent width (EW) of \hal\ is nearly independent of galaxy luminosity. { The lines show the 5th-, 25th-, 50th-, 75th-, and 95th-percentiles as a function of absolute magnitude.}}
\label{fig:ew_mag}
\end{center}
\end{figure}

First, we bin the quiescent red galaxy sample by the strength of their emission lines. 
The equivalent widths of the sample show very little dependence on the galaxy luminosity (Figure~\ref{fig:ew_mag}). To detect the weak emission lines such as \niitw\ and \oiiitw, we focus on galaxies with relatively strong line emission. Later, we will use galaxies without line emission as a control sample for stellar continuum subtraction.  

As shown by \cite{Yan06} and in the top left panel of Figure~\ref{fig:ew_vs_ew}, most of these galaxies follow a very narrow sequence in \oii\ EW vs. \hal\ EW plot. In terms of \oii/\hal\ ratio, most of the sample span a very narrow range in \oii/\hal\ ratio. We suspect that the small fraction of galaxies that have low \oii/\hal\ ratio are a slightly different population. Their low \oii/\hal\ ratios could be due to several different reasons. They may either have contamination from another ionization mechanism, or have a different abundance pattern, or more dust extinction. We exclude this population of galaxies by the same cut as we used in \cite{Yan06}. We keep only galaxies that have ${\rm EW}(\oii) > 5 {\rm EW}(\hal)-7$. This cut removes 3.3\% of galaxies. It removes most of the Seyferts in the red sequence which have high \oiii/\oii\ ratio (\oiii/\oii$>1$) compared to the majority of quiescent galaxies.

We then limit the sample to those with a median S/N per pixel greater than 15. This cut is chosen to maximize the S/N in the stacked spectra. This would remove 27.5\% of galaxies after the previous cuts.

Next, we select the strongest line emitting galaxies. We base our selection on equivalent width of lines.




We do not base the selection of strong-line galaxies on \hal\ EW alone, as that would bias the line ratio distribution of the sample. For example, selecting strong \hal\ galaxies without considering \nii\ may bias the sample towards low \nii/\hal\ galaxies. Therefore, we construct a 'Total EW' by combining the EW of all strong emission lines (\hal, \nii, \oii, \oiii, \sii) according to the median EW ratios among these lines. This population populates a narrow sequence in many of the EW vs. EW plot (Figure~\ref{fig:ew_vs_ew}). For example, in \oii\ EW vs. \hal\ EW plot, these galaxies populate a narrow sequence with a slope of approximately 5. Order the galaxies by 5EW(\oii)+EW(\hal) will sort the galaxies along that sequence. Similarly, we obtain the EW ratios between other lines and \hal, and we construct a total EW index based on the following formula.

\begin{eqnarray}
{\rm Total~EW~index} &= &{\rm EW}(\hal) + 1.03 {\rm EW}(\nii) \nonumber\\
& & +5.0 {\rm EW} (\oii) + 0.5 EW(\oiii) \nonumber\\
& & + EW(\sii) 
\end{eqnarray}

We select a sample of strong emission line galaxies by requiring this Total EW index to be above 75 percentile in each redshift bin with $\Delta z= 0.1$. 
This leads to a sample of 18,664 galaxies.

We will bin the strong line sample by their \nii/\oii\ and \nii/\hal\ line ratios. Therefore, we require that the fractional errors on \nii/\oii\ and \nii/\hal\ ratios to be better than 0.3 dex. This removes 21\% of the strong line sample.  We further require \oiii/\oii\ ratio to be less than 1. This removes 1.2\% of the remaining sample. This yields our final strong-line sample with a total of 14,645 galaxies.

\begin{figure}
\begin{center}
\includegraphics[width=0.45\textwidth]{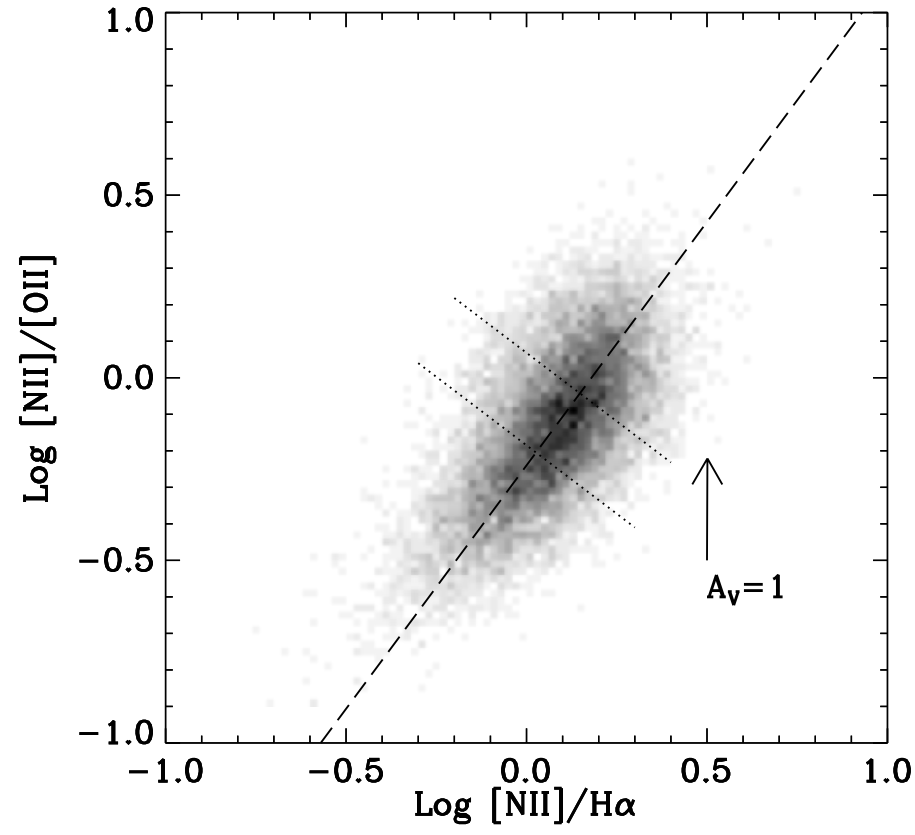}
\caption{Distribution of the selected strong-line quiescent red galaxies in \nii/\oii\ vs. \nii/\hal\ space. The gray scale indicates density of points. The dashed line is an approximation (not a fit) to the trend. The dotted lines mark our separation thresholds to split the sample into the high-\nii/\hal, mid-\nii/\hal, and low-\nii/\hal\ subsamples.}
\label{fig:n2o2_n2ha_den}
\end{center}
\end{figure}

With these cuts, the strong-lined quiescent red galaxies selected populated a relatively narrow locus in the \nii/\oii\ vs. \nii/\hal\ diagram as shown in Figure~\ref{fig:n2o2_n2ha_den}. \nii/\oii\ is an excellent metallicity indicator, because N and O have very similar ionization potentials for multiple ionization levels. Therefore, they always trace roughly the same spatial regions, no matter it is photoionization or collisional ionization. In addition, this line ratio is sensitive to metallicity because of two reasons. First, Nitrogen is a secondary element so N/O ratio goes up in high metallicity regime. Second, the two lines differ a lot in the transition energy, making it very sensitive to temperature variation which traces metallicity variation. Both factors increase \nii/\oii\ ratio as metallicity increases making this line ratio very sensitive to metallicity. 

The drawback with this line ratio is that it is quite sensitive to dust extinction. { However, this is of little consequence for our analysis of quiescent galaxies since the extinction vector goes vertically in Figure~\ref{fig:n2o2_n2ha_den}, which therefore differs strikingly from the general slope apparent in the data. As a result, such trend does not appear to be due to variations in dust extinction. At a fixed \nii/\hal\ ratio, the small observed variations in \nii/\oii\ imply that the extinction variations are relatively small among the selected quiescent galaxies.}

Gas with different metallicities can have different electron temperatures. It therefore makes more sense to separate the whole sample into different metallicity bins. We separate this sample into three bins according to their position in the \nii/\oii\ vs. \nii/\hal\ space, which should roughly trace metallicity variations. 

We find an approximation to the trend in this space and cut the sample into 3 equal portions along the dotted lines as shown in Figure~\ref{fig:n2o2_n2ha_den}. Below, we refer to the three subsamples as high-\nii/\hal, mid-\nii/\hal, and low-\nii/\hal\ subsamples.

\subsection{Constructing the matching zero-line subsamples}
  In order to measure the weak emission lines, we have to accurately subtract the stellar continuum underneath the lines. One of the best ways for stellar continuum subtraction is to use other galaxies without emission lines to build a stellar continuum template. Thus, we construct a sample of red quiescent galaxies without emission lines and select a matching subsample for each strong-line subsample. 
  
  We select those galaxies that center around zero EW in all the strong emission lines (\hal, \niibw, \siiw, \oiiibw, \hb\, and \oiiw). We draw a multi-dimensional ellipsoid centered around 0 (the origin) in the multi-dimensional space with each of the semi-axis equal to two times the median uncertainty in EW for each line. Again the selection is done in each redshift bin as the uncertainties change with redshift. This yields a sample of 13,353 galaxies in the zero-line sample. 
    
  Because there are weak but non-zero correlations between the emission-line strength and the stellar population, the high-\nii/\hal\ strong-line subsample and the low-\nii/\hal\ strong-line subsample may have different stellar continuum. To accurately reproduce the stellar continuum of each strong-line subsample, the galaxies making up the corresponding zero-line subsample needs to have the same stellar population and the same velocity dispersion. We experimented several different matching methods. The best continuum subtraction is achieved by matching the 3-d distribution of galaxies in $r_{0.1}$-band absolute magnitude, $D_n(4000)$, and stellar velocity dispersion space. 
  
In detail, we first choose the bin size in each dimension (each property to be matched) so that there are 20 bins between the 5-th and 95-th percentile points in that property for the zero-line sample. This corresponds to $\Delta M_{0.1r}=0.119$, $\Delta D_n(4000)=0.011$, and $\Delta \log V_{\rm disp}= 0.016$. 
For each strong-line subsample, we go through each bin in this 3-d parameter space and randomly draw twice as many galaxies in the zero-line sample belonging to that bin as there are in the given strong-line subsample for that bin. If a bin has fewer galaxies in the zero-line sample than there are in the strong-line subsample, then we discard all galaxies in that bin from both samples. If the zero-line sample has more than twice the number of the strong-line subsample, we do the random drawing without replacement. If the zero-line sample has more galaxies in that bin than the strong-line subsample but less than twice as many, we allow each galaxy in the zero-line sample to be selected at most twice. This way, we minimize repetition of spectra in constructing the corresponding zero-line subsample. 

\subsection{Stacking the Spectra}
  We stack the spectra of all galaxies in each strong-line subsample and each zero-line subsample. For the stacking, we first correct each spectrum for the Milky Way galactic extinction using the extinction map of \cite{SchlegelFD98} and the extinction law of \cite{O'Donnell94}. We then correct each spectrum by the correction vector derived by \cite{Yan11flux} (done in observed wavelength space), normalize it by the median flux in the window of 6000-6100\AA, and re-sample it to a common wavelength grid with the same logarithmic spacing. Finally, we sum together all spectra in a given sample and propagate the errors accordingly. 

\begin{figure*}
\begin{center}
\includegraphics[width=1.0\textwidth]{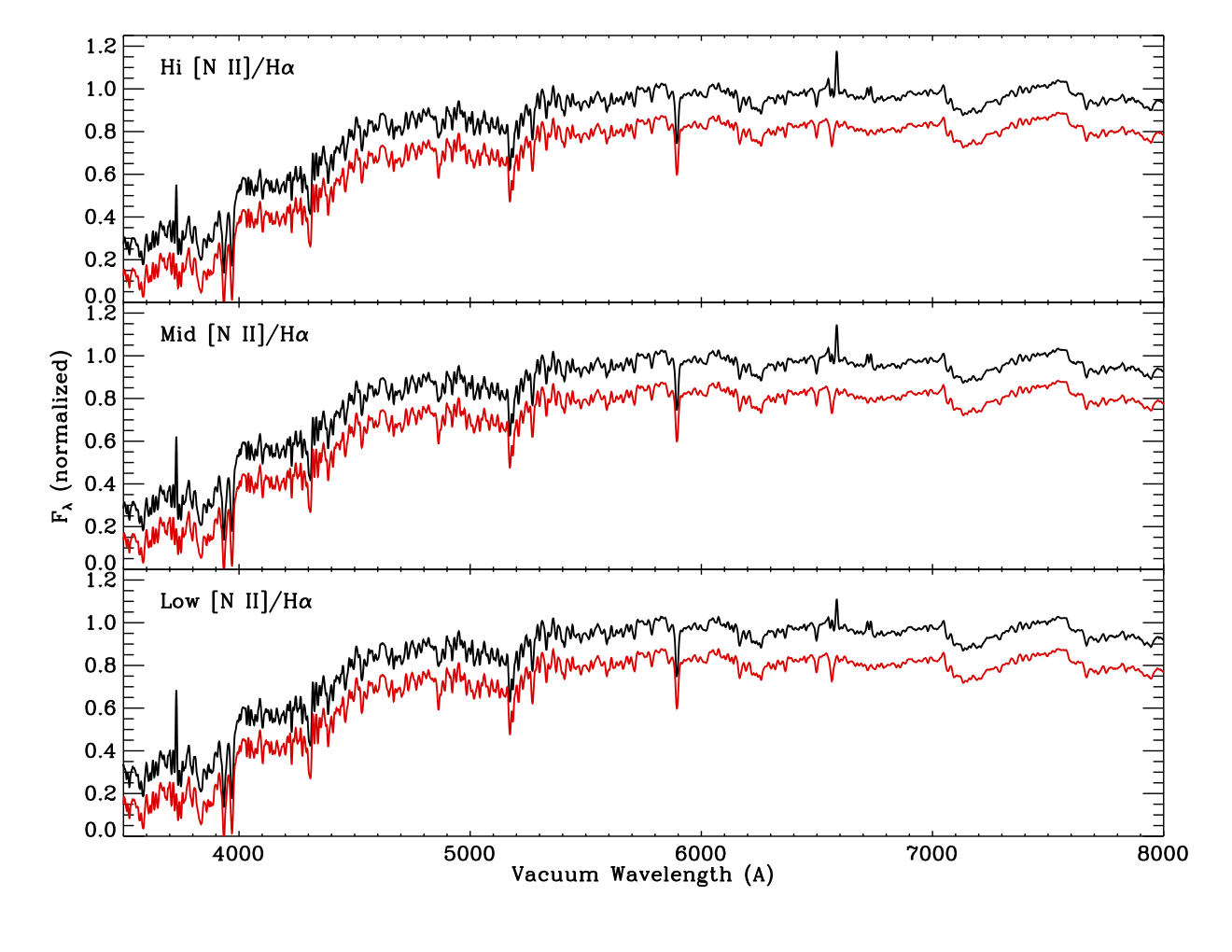}
\caption{Stacked spectra for the strong-line subsamples in black and for the zero-line subsamples in red. The three panels are for the three \nii/\hal\ bins, respectively. The spectra for the zero-line subsamples are offset vertically for clarity.}
\label{fig:coaddD_comparison}
\end{center}
\end{figure*}

 Figure~\ref{fig:coaddD_comparison} shows the stacked spectra for the high \nii/\hal\ sample, the medium \nii/\hal\ sample, and the low \nii/\hal\ sample. 

\begin{figure*}
\begin{center}
\includegraphics[width=1.0\textwidth]{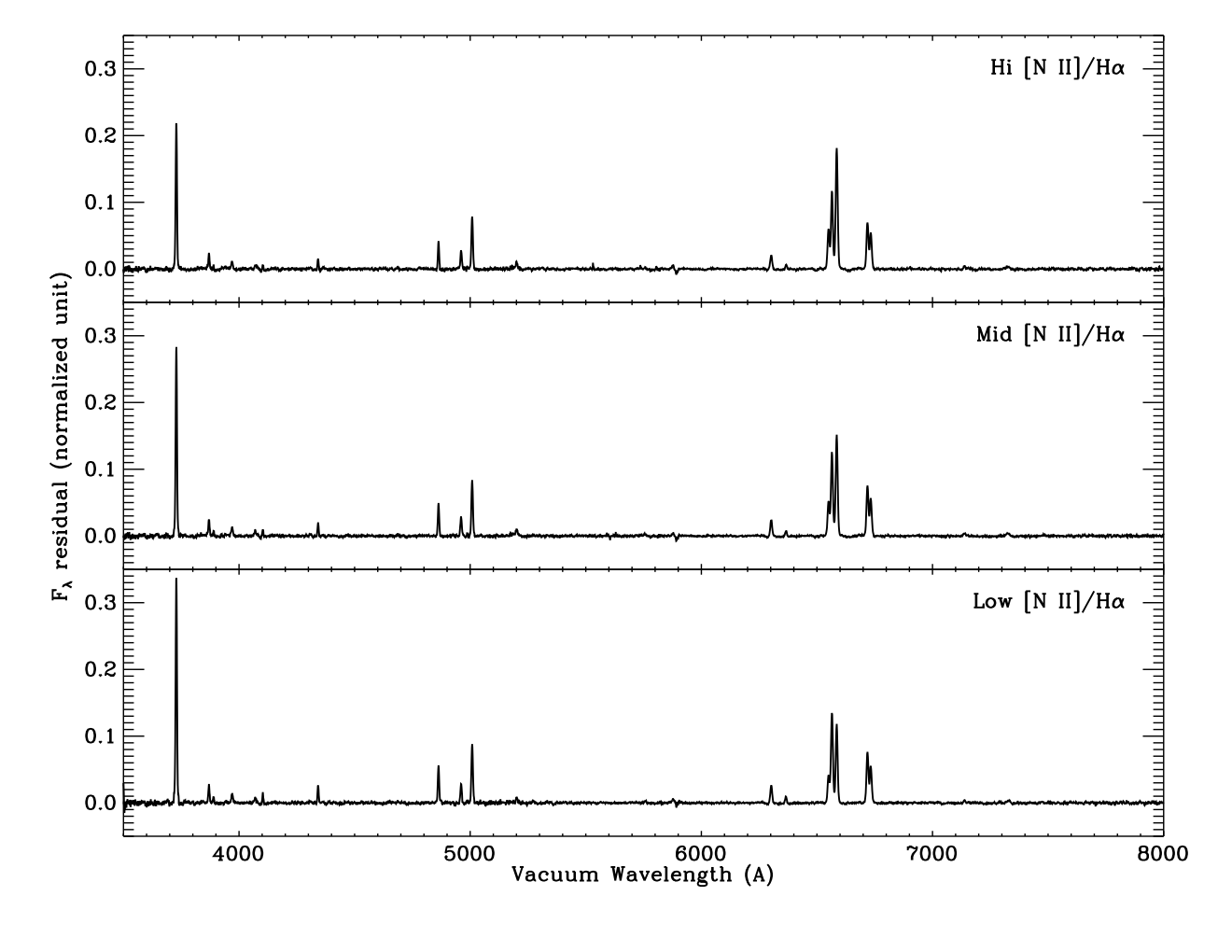}
\caption{Continuum-subtracted residual stack spectra for the three strong-line subsamples with different \nii/\hal\ ratios. }
\label{fig:overallresidual}
\end{center}
\end{figure*}

 To measure the emission lines, we need to subtract the continuum.  Although we have matched the zero-lined galaxies with the strong-lined galaxies before stacking their spectra, the resulting continuum still have a slightly different broadband shape at a few percent levels. This would introduce residual features in the continuum. To avoid this problem, we first divide the stack spectrum of each strong-line subsample by the stack spectrum of its corresponding zero-line subsample, then do a b-spline fit through the ratio spectrum using only the line-free regions and with a spacing of 400 pixels between break points. We then multiply the zero-line stack with this smooth ratio curve before subtracting it from the strong-lined stack. This is done separately for each strong-line subsample. The resulting residual spectrum are shown in Figure~\ref{fig:overallresidual}. One can see the residuals are flat to much better than 1\%.  

\subsection{Emission line measurements in the stacked spectra}

\begin{figure*}
\begin{center}
\includegraphics[width=0.49\textwidth]{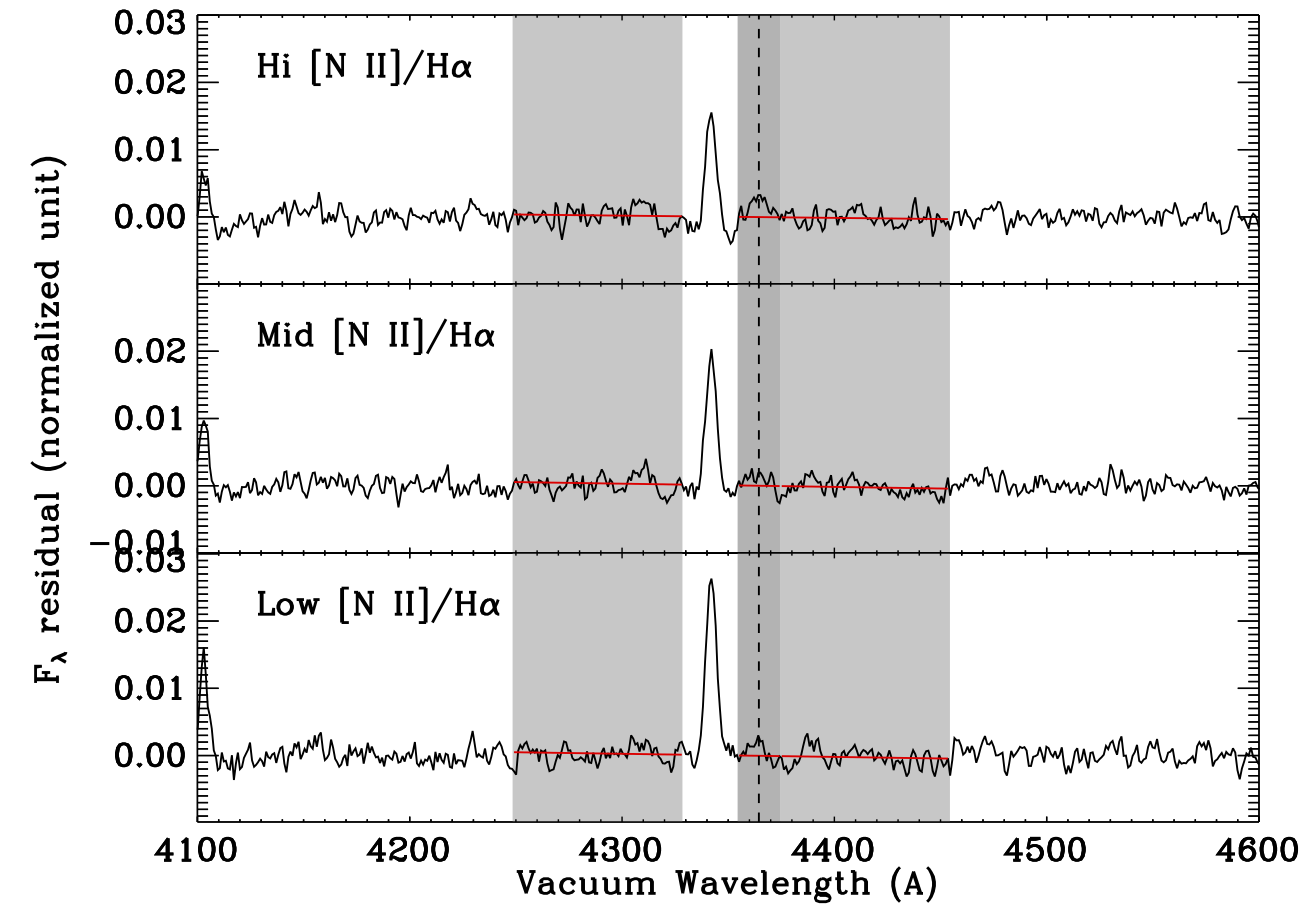}
\includegraphics[width=0.49\textwidth]{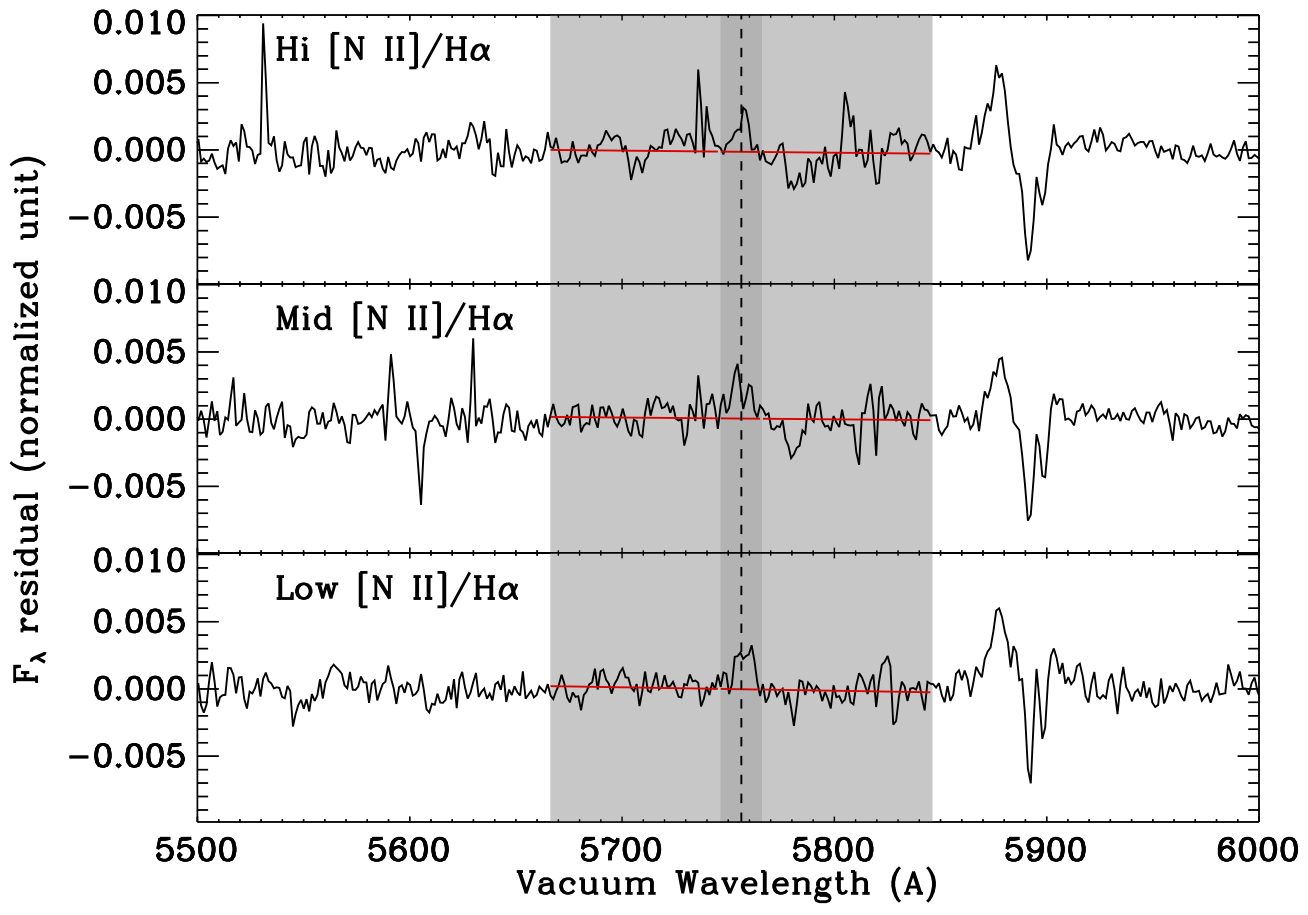}
\includegraphics[width=0.49\textwidth]{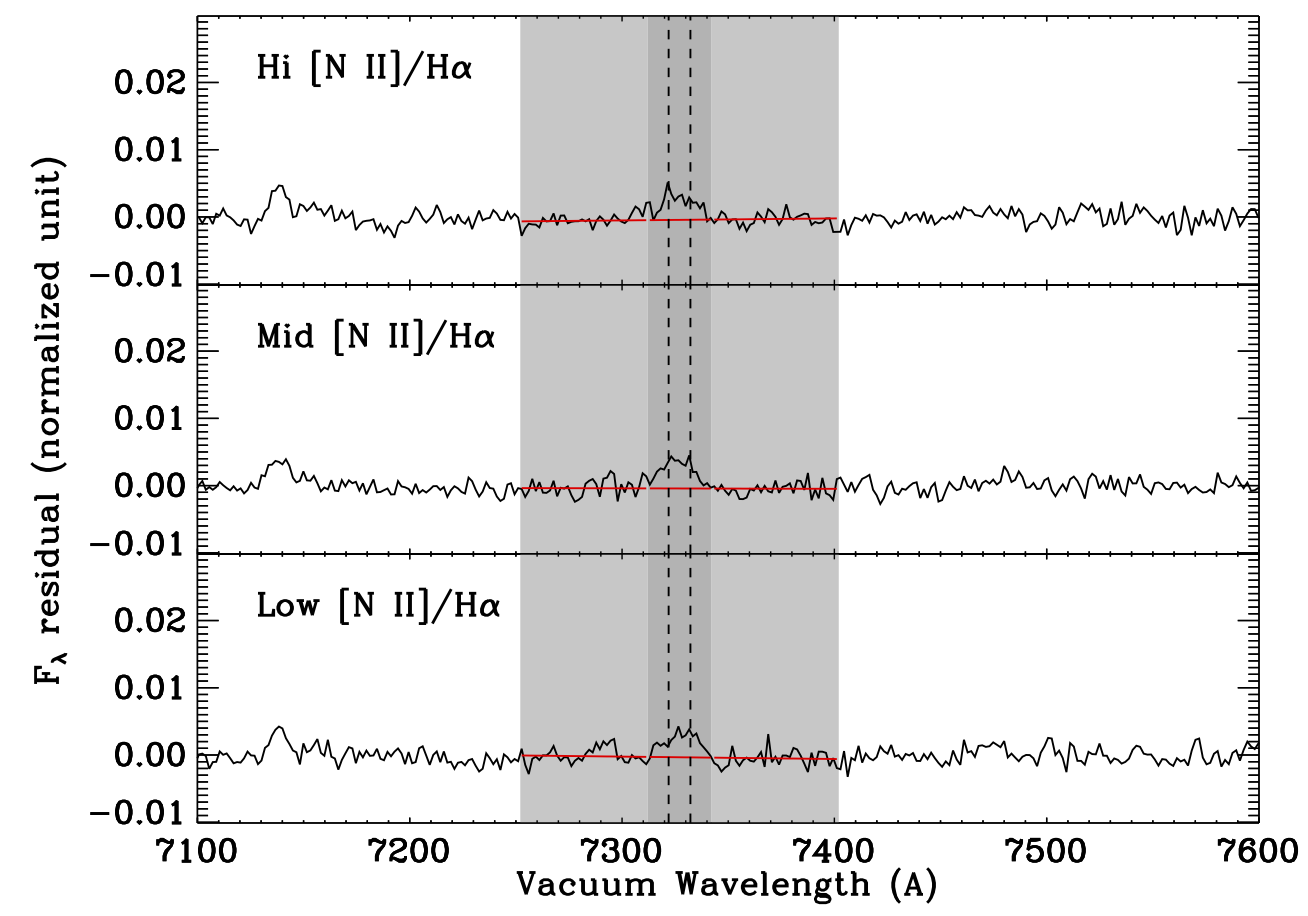}
\includegraphics[width=0.49\textwidth]{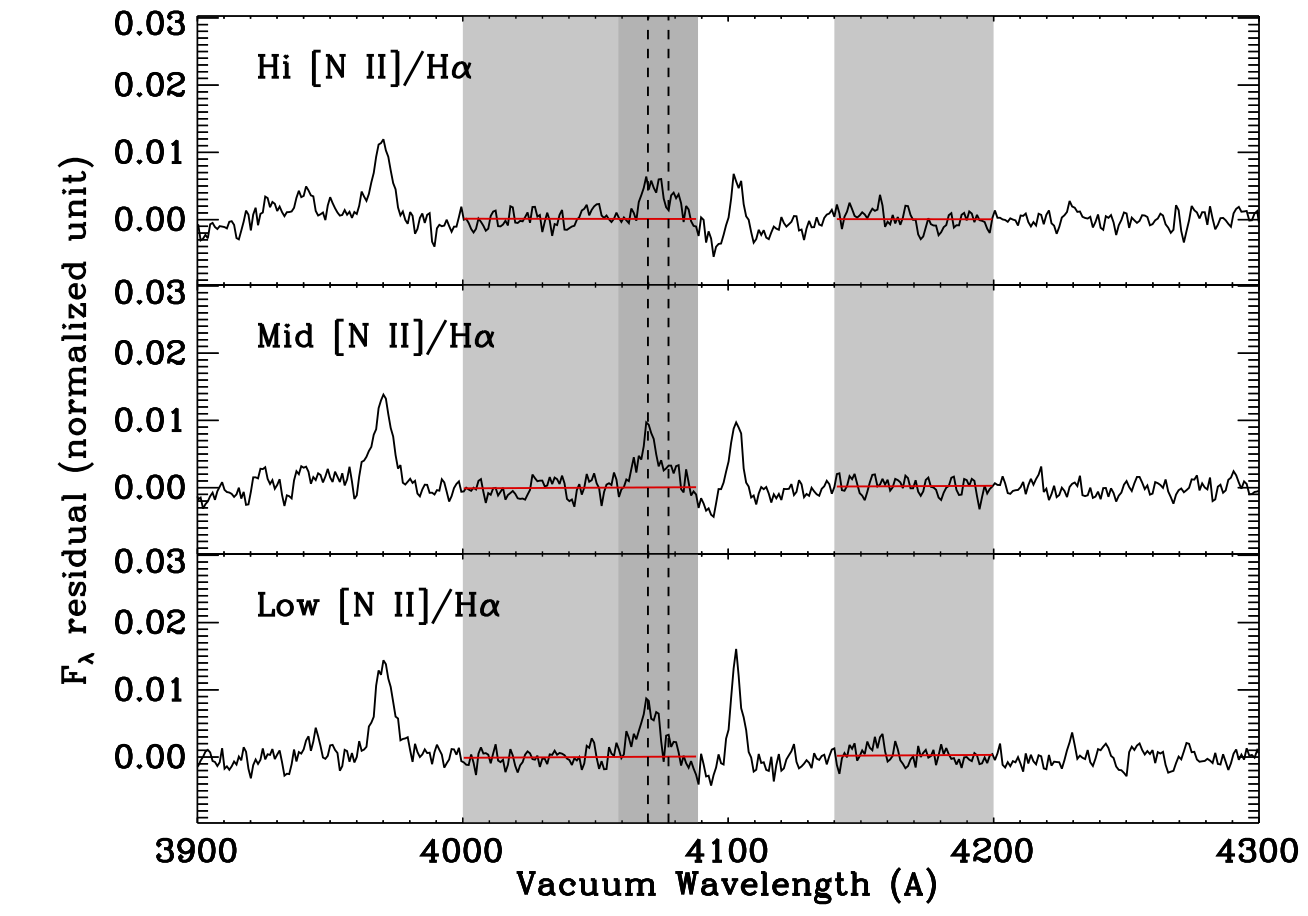}
\caption{Zoom-in of the continuum-subtracted stack spectra around \oiiitw (top left), \niitw (top right), \oiitw (bottom left), and \siitw (bottom right) for the three strong-lined subsamples. The light grey zones indicate the wavelength windows used to determine the residual continua, which are fit by straight lines (red solid lines). The dark gray zones indicate the windows over which the line flux is integrated. The expected line centers are shown by the vertical dashed lines.}
\label{fig:coronalzoomin}
\end{center}
\end{figure*}

 We measure the emission lines in the residual spectra using similar algorithm as given by \cite{Yan06}. For each emission line, we measure the residual continuum level in two sidebands as given in Table~\ref{tab:linedef} and \ref{tab:weaklinedef}. We fit a linear function through the two sidebands and subtract it from the spectra before measuring the line flux. We sum the residual flux in the line window as the resulting line flux. { For all the Balmer lines, we adopt the Gaussian-fitted line flux rather than summed flux as visual inspection suggests the Gaussian-fitting results in better fits to the continuum levels for the weaker Balmer lines.} 

\begin{table*}
\begin{tabular}{llll}
\hline\hline
Line & Center window (\AA) & Left sideband(\AA) & Right sideband (\AA)\\ \hline
\hz & 3884.17---3896.17 & 3835.00---3860.00 & 3896.18---3921.18 \\
\heps & 3963.20---3979.20 & 3952.20---3960.20 & 3979.20---3987.20 \\
\siitw & 4058.625---4088.625 & 4000---4060 & 4140---4200\\
\hd & 4094.892---4110.892& 4086.892---4094.892 & 4110.892---4118.892\\
\hg & 4334.692---4348.692& 4326.692---4334.692 & 4348.692---4356.692 \\
\oiiitw & 4354.435---4374.435 & 4248.435---4328.435 & 4374.435---4454.435 \\
\niitw & 5746.2---5766.2 & 5666.2---5746.2 & 5766.2---5846.2 \\
\oiitw & 7312.08---7342.08 & 7252.08---7312.08 & 7342.08---7402.08\\
\hline\hline
\end{tabular}
\caption{Definition of windows for weak emission line measurements in the continuum-subtracted stack spectra.}
\label{tab:weaklinedef}
\end{table*}

 The fluctuations in line-free regions of the residual spectra show the level of uncertainty associated with our continuum subtraction technique. For the strong forbidden lines, the uncertainty in the summed line flux is propagated from the uncertainties in the stacked spectra. For the weaker coronal lines, we adopt a more conservative estimate that would take into account any systematics resulting from imperfect continuum subtraction. We simulate the flux measurements with a sliding boxcar placed in the sidebands with the same width as the line window. We take the root-mean-square of about 100 to 160 integrated measurements of fluxes in such regions without any emission lines as the uncertainty of the line measurement. We show in Table \ref{tab:lines} our measurements for all emission lines in the continuum-subtracted spectra. { The first three columns gave the raw values relative to Hbeta flux, without correcting for intrinsic dust extinction. The second set of three columns show the values after extinction correction according to the \hal/\hb\ ratio. However, as we will discuss below, we have reasons to suspect that these extinction correction can be unreliable. Thus, we do not recommend using these extinction-corrected values. }
 
Figure~\ref{fig:coronalzoomin} shows the residual spectra around the four temperature-sensitive coronal lines, which are the key measurements for this paper. In the bottom panels, one can see that the \oiitw\ and \siitw\ lines are significantly detected in all three stacked spectra. The bump around 7290\AA\ in the low-\nii/\hal\ stack could be due to \caiiaw. The \niitw\ line is detected in the mid- and low-\nii/\hal\ stacks. There is a bump around the wavelength of \niitw\ in the high-\nii/\hal\ stack but it is of a similar level as other features that are due to noise. According to our simulated flux measurements in the sidebands as described above, it is less than 3$\sigma$. The features to the right of the shaded region are \heiw\ in emission and \naiw\ in absorption. 
{ For \oiiitw, the bumps in all three spectra around that wavelength do not look significant. This spectral region could be affected by an imperfect subtraction of the \hg\ absorption feature present in the stellar continuum. Only in the high \nii/\hal\ stack is the \oiiitw\ marginally detected, at a 3$\sigma$ level.}
 
\begin{table*}
\begin{tabular}{ m{4cm} |ccc|ccc}
\hline\hline
Lines & Hi-\nii/\hal\ & Mid-\nii/\hal\ & Low-\nii/\hal\ & Hi-\nii/\hal\ & Mid-\nii/\hal\ & Low-\nii/\hal\  \\ 
 & \multicolumn{3}{c|}{Raw measurements} & \multicolumn{3}{c}{Extinction-corrected (not recommended)}\\
\hline

\hal\ & $ 439.2\pm  8.0$& $ 393.6\pm  6.1$& $ 390.6\pm  5.7$& $ 287.0\pm  5.2$& $ 287.0\pm  4.4$& $ 287.0\pm  4.2$\\
\hb\ &  $100$ & $100$ & $100$ &  $100$ & $100$ & $100$\\ 
\oiiibw\ & $ 229.0\pm  4.5$& $ 200.9\pm  3.4$& $ 191.7\pm  3.0$& $ 217.0\pm  4.2$& $ 193.0\pm  3.2$& $ 184.3\pm  2.9$\\
\niibw\ & $ 687.7\pm 12.3$& $ 459.4\pm  7.0$& $ 312.1\pm  4.6$& $ 447.6\pm  8.0$& $ 334.0\pm  5.1$& $ 228.7\pm  3.4$\\
\siiw\ & $ 466.9\pm  8.5$& $ 394.8\pm  6.2$& $ 363.1\pm  5.4$& $ 296.0\pm  5.4$& $ 281.4\pm  4.4$& $ 261.0\pm  3.9$\\
\oiiw\ & $ 583.6\pm 10.6$& $ 620.9\pm  9.5$& $ 686.0\pm  9.8$& $ 925.2\pm 16.8$& $ 874.2\pm 13.4$& $ 957.9\pm 13.7$\\
\oiiitw\ & $  10.9\pm  3.1$& $   4.1\pm  3.1$& $   2.9\pm  2.8$& $  13.6\pm  3.8$& $   4.8\pm  3.6$& $   3.4\pm  3.2$\\
\niitw\ & $   7.7\pm  5.2$& $   9.3\pm  2.3$& $   8.2\pm  1.7$& $   5.9\pm  4.0$& $   7.7\pm  1.9$& $   6.7\pm  1.4$\\
\siitw\ & $  24.3\pm  3.6$& $  25.4\pm  2.4$& $  19.7\pm  2.1$& $  34.1\pm  5.0$& $  32.7\pm  3.0$& $  25.3\pm  2.7$\\
\oiitw+\caiibw\ & $  24.6\pm  2.8$& $  23.2\pm  2.3$& $  18.9\pm  2.1$& $  13.9\pm  1.6$& $  15.2\pm  1.5$& $  12.5\pm  1.4$\\
\oiitw\ & \multirow{2}{4em}{$24.6\pm  4.0$}& \multirow{2}{4em}{$20.6\pm  2.5$}& \multirow{2}{4em}{$  14.0\pm  4.5$}&   \multirow{2}{4em}{$13.9\pm  2.3$}& \multirow{2}{4em}{$  13.5\pm  1.7$}&\multirow{2}{4em}{ $   9.3\pm  3.0$}\\
 (corrected for \caiibw)& & & & & &\\\hline
\siibw/\siiaw\ & $0.778\pm0.007$ & $0.741\pm0.007$ & $0.719\pm0.009$ & $0.776\pm0.007$ & $0.740\pm0.007$ & $0.717\pm0.009$ \\
\hb\ Flux &  $ 0.276\pm0.005$& $ 0.340\pm0.005$& $ 0.372\pm0.005$& $ 0.382\pm0.007$& $ 0.433\pm0.006$& $ 0.471\pm0.006$\\
\hline\hline
\end{tabular}
\caption{{ Measurements of strong and weak lines in the continuum-subtracted stacked spectra. The measurements are given relative to the flux of the \hb\ line, which is given in the bottom row. The left three columns show the raw ratios. The right three columns show the ratios after correcting for intrinsic dust extinction according to \hal/\hb\ ratios. As discussed in the text, this extinction correction may not be reliable as it could lead to unphysical ratios between other lines.} The unit of the \hb\ line fluxes is Angstrom multiplied with the flux density between 6000\AA\ and 6100\AA, since each spectrum is normalized by the median flux in the window 6000-6100\AA\ before being stacked.}
\label{tab:lines}
\end{table*}


%

\section{Densities and Temperatures of the Low-ionization Gas}\label{sec:measurements}

\subsection{Balmer Decrements and Extinction Correction}
\begin{figure}
\begin{center}
\includegraphics[width=0.5\textwidth]{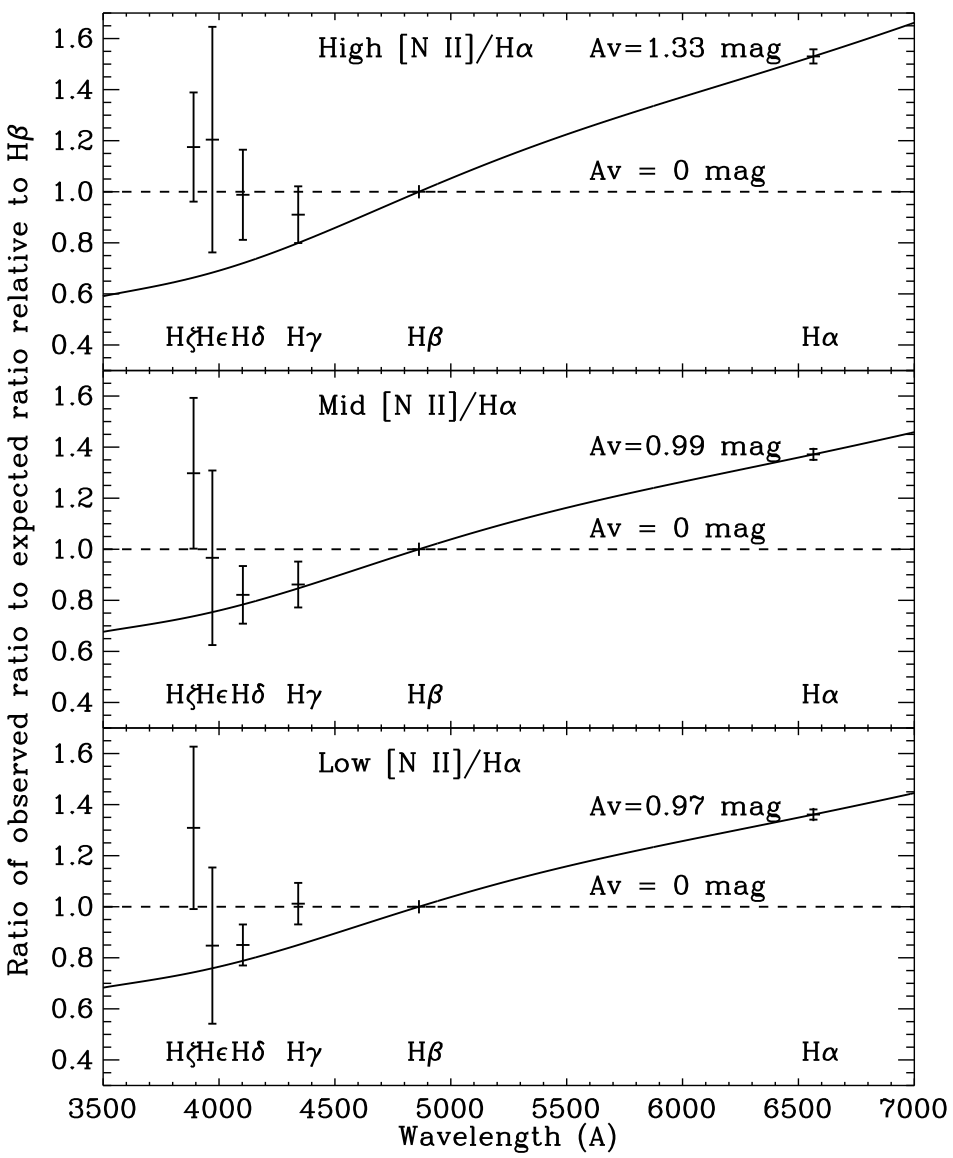}
\caption{We show the ratio between the observed Balmer-line to \hb\ ratio and the theoretical case B Balmer-line to \hb\ ratio. The error bars indicate $\pm1\sigma$ uncertainties. The three panels are for the three \nii/\oii\ subsamples. If there was no dust extinction, we expect the points to lie on the horizontal dashed line. The curves indicate the expected trend under the level of dust extinction constrained from the \hal/\hb\ ratio. }
\label{fig:BalmerDecrements}
\end{center}
\end{figure}

The average level of dust extinction in each subsample can be measured by comparing the observed Balmer decrements with the theoretical values expected under Case B recombinations. Multiple Balmer lines, from $H\alpha$ to $H\zeta$, are detectable in the stacked, continuum-subtracted  spectra. Therefore, we have multiple decrements we can use to evaluate extinction. Figure~\ref{fig:BalmerDecrements} shows how the ratio between various Balmer lines and \hb\ compare with the Case B expectations in the low-density limit, { assuming $T=10,000$K}. Here we use the Gaussian-fitted line fluxes rather than the summed fluxes to make the measurements less affected by residual Balmer absorption. H$\epsilon$ ($\lambda$3970\AA)overlaps with the \neiiibw\ line, which is part of a doublet with the \neiiiw\ line with a fixed ratio. We measure the \neiiiw\ and subtract 31\% of its flux from the Gaussian fit at the wavelength of H$\epsilon$. 

The curves in Fig.~\ref{fig:BalmerDecrements} show the expected trend under a level of dust extinction constrained by the \hal/\hb\ ratio. The higher order Balmer lines tend to show a smaller level of extinction or zero extinction, but are broadly consistent within 1-2$\sigma$ with the curve determined through \hal/\hb\ ratio. Given the high S/N on the \hal\ and \hb\ lines, one would think the \hal/\hb\ ratio would give a reliable extinction correction. However, the reality is more complex. As we will see below, the combination of the two temperature-sensitive line ratios, \oiitw/\oiiw\ and \siitw/\siiw, can also provide an extinction estimate (also see \citealt{Dopita82}). They yield much smaller extinction estimates than that given by \hal/\hb\ ratio, with upper limits in $A_{\rm v}$ to be 0.48, 0.14, an 0.0 mag for the three subsamples. Thus, we decide not to correct the line ratios for extinction. We will come back and discuss this point below. 


\subsection{ Density Derivation}

{ We compute the density of the gas using the \siibw/\siiaw\ line ratio from the stacked spectra. This line ratio is a good density indicator and it is insensitive to temperature. We compute the density using the {\it nebular.temden} routine \citep{ShawD95} in PyRAF, which is based on the 5-level atom program written by \cite{deRobertis87}. Assuming $T=10,000$K, we obtain the mean density values for the three subsamples, listed in Table~\ref{tab:tempraw}. The mean electron density for the high-, mid-, and low-\nii/\hal\ samples are $136\pm14$ cm$^{-3}$, $71\pm12$ cm$^{-3}$ and $34\pm15$ cm$^{-3}$, respectively.
The 1-$\sigma$ uncertainties in density are computed according to the uncertainty of the line ratio measurements. This uncertainty is the uncertainty of the mean density among all galaxies in each subsample. It does not indicate the uncertainty for an individual galaxy nor the range of densities among galaxies. 

This line ratio is also measurable in individual galaxies with significant \sii\ detection. If we limit to only those strong-line quiescent galaxies with \sii\ detected above 10$\sigma$, we find the RMS scatter in the \siibw/\siiaw\ ratio is 0.11 with a median ratio of 0.79. This scatter still includes significant measurement noise as the median \sii\ EW for them is only 4.2\AA. But we can see the density of the \sii-emitting gas in these galaxies is always in the low density regime. 

The fact that the line-emitting gas in these quiescent galaxies { takes place} in the low density regime { with respect to the optical \nii\ and \oiii\ lines} means that their { strengths} are not affected by collisional deexcitation. Since we expect O$^{++}$ to lie at a higher temperature than S$^+$, we expect it to be emitted from an even lower density plasma than that of O$^+$ and S$^+$. Thus, it is unlikely for density stratification to affect the use of \oiiitw/\oiiiw\ as temperature indicators, or the use of other coronal-to-strong line ratios in this paper. We are safe to use them to measure temperatures. }

{ As we will discuss below, the implication of this density measurement for shock models is that a much lower pre-shock density than what we infer here for \sii\ is required to match the \sii\ and \oii\ data.}

\subsection{Temperature Derivation}

Different ions trace regions with different ionization in an ionized cloud. The \oiii\ lines trace the more highly ionized regions which have more O$^{++}$. The \oii, \nii, and \sii\ lines trace the partially ionized regions, as O$^0$, N$^0$, and S$^0$ have an ionization potential similar to H$^0$. In the following discussion, we should keep in mind the temperature measured from \oiii\ could be different from the temperatures measured from \oii, \nii, and \sii, as they trace different emission line regions. 


We derived the temperature for the gas using several pairs of line ratios. These are also computed using the {\it nebular.temden} routine mentioned above. 
Among the four temperature indicators we use, \oiiiw/\oiiitw\ and \niiw/\niitw\ have very little secondary dependence on density, while \siiw/\siitw\ and \oiiw/\oiitw\ have slightly higher secondary dependence on density, where a factor of 10 difference in the assumed density would induce a 5-10\% change in the derived temperature. { We consistently assumed the density inferred from the \siiw\ doublet in the evaluation of all these temperature indicators. }

We list the temperatures derived and the 1-$\sigma$ uncertainties in Table~\ref{tab:tempraw}.{ We list the values before and after extinction correction. However, we recommend using the set without extinction correction as there are reasons to believe the Balmer-decrements are yielding unreliable extinction measurements.}

\begin{table*}
\begin{tabular}{l|ccc|ccc}
\hline\hline
Indicators & Hi-\nii/\hal\ & Mid-\nii/\hal\ & Low-\nii/\hal\ &Hi-\nii/\hal\ & Mid-\nii/\hal\ & Low-\nii/\hal\ \\ 
(Unit: cm$^{-3}$ for $n$ and $10^4$K for $T$)& \multicolumn{3}{c|}{Before extinction-correction} & \multicolumn{3}{c}{Extinction-corrected (not recommended)}\\ \hline
$n_e from {\siibw/\siiaw}$   &  $136\pm14$ &  $71\pm12$ & $34\pm15$ &  $133\pm13$ &  $70\pm12$ & $31\pm15$\\
$T_{\oiiitw/\oiiibw}$ &  $2.59^{+0.93}_{-0.42}$ & $<1.93$ & $<1.87$ & $3.29^{+1.70}_{-0.66}$ & $<2.17 $ & $ <2.07 $ \\
$T_{\niitw/\niibw}$  & $<1.05$ &$1.20^{+0.19}_{-0.12}$ & $1.37^{+0.20}_{-0.13}$ & $<1.12$	&$1.27^{+0.23}_{-0.13}$ & $1.46^{+0.24}_{-0.14}$\\
$T_{\siitw/\siiw}$& $0.72^{+0.07}_{-0.05}$& $0.84^{+0.06}_{-0.05}$ &$0.78^{+0.05}_{-0.05}$ & $1.29^{+0.27}_{-0.16}$& $1.35^{+0.17}_{-0.12}$ &$1.16^{+0.14}_{-0.10}$\\
$T_{\oiitw/\oiiw}$&$1.71^{+0.3}_{-0.23}$ & $1.48_{-0.14}^{+0.20}$ & $1.10^{+0.35}_{-0.16}$ &$0.87^{+0.09}_{-0.06}$ & $0.92_{-0.06}^{+0.07}$ & $0.77^{+0.15}_{-0.08}$\\
\hline\hline
\end{tabular}
\caption{Electron density and temperature measurements derived using line ratios before (left three columns) and after (right three columns) correcting for intrinsic dust extinction. We do not recommend using the second set of values as the extinction derived from the Balmer decrement could be unreliable. We discuss this in detail in Sections \ref{sec:measurements} and \ref{sec:constraints}.}
\label{tab:tempraw}
\end{table*}


When interpreting the line ratios, we should keep in mind { that the line ratios in the stacked spectra is a weighted average. }
Spectra stacking is done after normalizing each spectrum at 6000-6100A. Therefore, the line ratio in the stacked spectra is a weighted average with the weight proportional to the ratio between the denominator line and the 6000-6100A continuum. This can be shown by the derivation below, where we use $F_{\rm A}$ and $F_{\rm B}$ to denote the flux of two emission lines in the stacked spectra. The flux in individual spectra are denoted as $F_{i{\rm A}}$ and $F_{i{\rm B}}$. The continuum level between 6000-6100A is denoted as $C_i$. 
\begin{align}
({F_{\rm A}\over F_{\rm B}})_{\rm stack} &= {\sum_i  {F_{i{\rm A}}\over C_i} \over \sum_i {F_{i{\rm B}}\over C_i}} \\
&= {\sum_i {F_{i{\rm A}} \over F_{i{\rm B}}} {F_{i{\rm B}} \over C_i} \over \sum_i {F_{i{\rm B}}\over C_i}}
\end{align}For example, the \hal/\hb\ ratio in the stack is effectively a weighted average of the Balmer decrement, with the weight being the ratio between \hb\ and 6000-6100A continuum. Galaxies with stronger lines will have a higher weight and galaxies with less extinction will have a higher weight.

This means that the various line ratios are using slightly different weighting schemes to compute the weighted average. This could potentially introduce small inconsistencies between line ratios when we compare them to models. However, the galaxies in our sample have fairly homogeneous continuum shape: the RMS variation in the ratios between the continuum around \oii\ and that around \hal\ is only 10.6\%. The emission line EWs also have very small dynamic ranges: RMS scatter in the logarithm of EWs is only 0.2 dex. Thus, any effect this effective weighting has on the final line ratio is expected to be small and does not affect our main conclusions.


\begin{enumerate}
\item Concerning the \oiiiw/\oiiitw\ ratio. We only detect significant \oiiitw\ emission in the high-\nii/\hal\ subsample, at a 3.5$\sigma$ significance. This yields a temperature of $2.59^{+0.93}_{-0.42}\times10^4$K without extinction correction. 
In the mid-\nii/\hal\ and low-\nii/\hal\ subsamples, the \oiiitw\ lines are undetected (below 2$\sigma$), indicating the temperatures are below 1.93$\times10^4$K and  1.87$\times10^4$K (2$\sigma$ upper limits), respectively.  The overall trend seems to be that the temperature in the O$^{++}$ zone is getting lower as metallicity decreases, which is inconsistent with our general expectation of higher temperature in lower metallicity gas.

Our measurement of \oiiitw\ could potentially be contaminated by the \feiibw\ line, which is very close to 4363 in wavelength. The signal-to-noise ratio of the \oiiitw\ line does not allow a reliable decomposition. In principle, we could estimate the strength of the line based on the strength of \feiiaw, which originates from the same upper level as \feiibw. { However, as we do not see any visible bump at 4288\AA\ in the residual spectrum (see Fig.~\ref{fig:coronalzoomin}), we do not attempt to make such correction. In order to evaluate the potential impact of the \feiibw\ line, we included it in the model calculation and compared it directly to the data.} The ratio of \feiibw/\oiiitw\ is strongly dependent on temperature, and decreases with increasing temperature. Given our temperature measurement based on \sii\ and  \nii, the contamination of \feiibw\ is expected to be less than 2\% of \oiiitw, in the photoionization model. The shock models and turbulent mixing models are expected to produce even higher temperatures, thus we expect \feiibw\ to have little impact on the \oiiitw\ line measurements.

We also note that, the marginally-detected \oiiitw\ line in the high-\nii/\hal\ subsample is about 30\% wider than the \oiiiw\ line. This could be due to either noise or that the line has significant contributions from gas with a much higher velocity dispersion that is presumably much hotter. { This could weaken our temperature measurement inferred from \oiiitw. We cannot rule out either that the wider \oiiitw\ profile may originate from a much denser region around the nuclei of these galaxies. }
We warn the reader that our marginal detection of that line is not { fully} robust. It should be emphasized, however, that our final conclusions do not depend on it.

\item Concerning the \niiw/\niitw\ ratio.  We detect the \niitw\ emission line significantly in the mid-\nii/\hal\ and low-\nii/\hal\ subsamples, at 4.0$\sigma$, and $4.9\sigma$ respectively, indicating temperatures in the range between  $1.2\times10^4$K and $1.4\times10^4$K. We do not detect it significantly in the high-\nii/\hal\ subsample indicating a $2\sigma$ upper limit in temperature of $1.05\times10^4$K. The overall trend is that the temperature in the N$^+$ zone gets lower as metallicity increases, consistent with our general expectations. This is the opposite trend compared to that in the O$^{++}$ zone.

Both \niiw/\niitw\ and \oiiiw/\oiiitw\ ratios are insensitive to extinction, as the lines involved are not too far apart from each other. Extinction corrections would not have changed significantly the temperatures inferred from our \nii\ and \oiii\ estimators (Table~\ref{tab:tempraw}).

\item Concerning the \siiw/\siitw\ and\\ \oiiw/\oiitw\ ratios. First, the \oiitw\ is actually a quadruplet. Second, it could be contaminated by the \caiibw\ line. A correction can be performed based on the strength of the \caiiaw\ line, which is seen in the low-\nii/\hal\ spectrum of the bottom-left panel of Figure~\ref{fig:coronalzoomin}. In both the photoionization and the shock models, the \caiibw\ is always about 68-70\% of the \caiiaw\ line. Hence, we have measured the \caiiaw\ line and subtracted 69\% of it from our \oiitw\ measurements. The corrected value constitutes our adopted \oiitw\ measurements (see Table~\ref{tab:lines}).  

Interestingly, we detect the \oii\ and \sii\ coronal lines at a significant level.  The temperature inferred for the S$^+$ zone without extinction correction is around 8000K.
The temperature inferred for the O$^+$ zone without extinction correction is around 11,000-17,000K.
Note that these line ratios are very sensitive to the extinction correction. If we applied the extinction correction derived from Balmer decrements, it would reverse the temperature relationship between the two zones, making the S$^+$ zone { apparently} hotter than the O$^+$ zone. { This would be unphysical as it would contradict photoionization calculations in which the S$^+$ zone is either at a similar temperature as the O$^+$ zone, or cooler because it extends deeper into the photoionized slab, in the transition zone towards neutral gas where the heating rate has gone much lower. The ionization of Sulphur is maintained by photons of energy lower than those absorbed by H$^0$. An even stronger argumnent could be made with shock models.}


\end{enumerate}

\section{Constraints on the Ionization Mechanisms}\label{sec:constraints}
In this section, we compare our measurements of temperature sensitive line ratios with the predictions from theoretical calculations for three different mechanisms: photoionization by hot evolved stars, shocks, and turbulent mixing. 

\begin{enumerate}
\item{Photoionization models: }
We model the line ratios produced by photoionization by hot evolved stars using CLOUDY version { v17.00}, last described by \cite{Ferland17}. The input ionizing spectrum is a 13 Gyr old simple stellar population spectrum with solar metallicity, computed by \cite{BC03}. { We also include cosmic radio to X-ray background { radiation} in the simulation,  as well as the galactic background cosmic rays, using the default values provided by Cloudy.} The ionization parameter ($\log U$) spans from -4.5 to -2. The metallicity spans $-1.2 < [O/H]< 0.6$ where [O/H]$ = \log (O/H) / \log (O/H)_\odot$. The solar $12+\log(O/H)$ is assumed to be 8.69 \citep{GrevesseASS10}. The abundance pattern is assumed to follow the solar abundance pattern except for Nitrogen. The N/O ratio is assumed to increase with the O/H ratio, as Nitrogen has a secondary nucleosynthesis contribution when O/H is high. We adopt the fitting formula given by \cite{VilaCostasE93} as shown in the following equation.
\begin{equation}
\log (N/O) = \log (0.034 + 120 (O/H))
\end{equation}

{ For Carbon abundance, we follow \cite{Dopita13} and assume it is always 0.6 dex higher than Nitrogen.  We use the default dust depletion factors in CLOUDY. Oxygen is depleted by 0.4 dex and Carbon is depleted by 0.6 dex, while Nitrogen is not depleted. }

{ The simulation assumes constant gas pressure, with an initial electron density of 100 cm$^{-3}$, and an open geometry. This assumption results in \siiw\ doublet line ratios consistent with what we observe. We have also verified that changing the density to 10 cm$^{-3}$ would hardly change the key lines ratios we use in this paper. We include attenuation by dust and photoelectric heating by dust in the ionized zone, adopting the Milky Way ISM grain type and size distribution.} 

{The model calculation for \oiitw\ includes all 4 lines in the quadruplet: 7318\AA, 7319\AA, 7329\AA, 7330\AA. The computation of all forbidden lines also include any potential recombination contribution and charge transfer contribution to the lines.} 

\item{Shock Models: }
The shock models come from \cite{Allen08} who presented a series of fully radiative fast shock models computed with the MAPPINGS III code, spanning a range of shock velocity, density, and magnetic field parameter ($B/n^{1/2}$). { The model data are available as part of the software ``IDL Tool for Emission-line Ratio Analysis (ITERA)''\citep{GrovesA10,GrovesA13}\footnote{https://github.com/astrobrent/itera}. }

{ In shock models, the post-shock cooling zone can have a much higher density than the pre-shock density.
When a shock passes through the gas, the density of the gas would be compressed by a factor of 4. It would then be compressed even further when the post-shock gas cools. The S+ zone only occurs when the gas has { sufficiently cooled and combined,} and thus corresponds to a much higher density than the pre-shock gas. To fit the data, the models first need to satisfy the measured density that is in the low density regime with \siibw/\siiaw having a mean ranging from 0.72 to 0.78 among the three subsamples. The RMS scatter of this ratio among individual galaxies is about 0.12 { if we consider only the \sii\ fluxes that reach $10\sigma$ or higher. Hence, we will require all models to produce a \siibw/\siiaw\ ratio less than 0.9. Among the shock models available, all those with a pre-shock density of either n=0.01 or 0.1 cm$^{-3}$ turn out to comply with our \siibw/\siiaw\ limit ratio (extending from the theoretical minimum value of 0.68 up to 0.9). All shock models with a pre-shock density of n=1 cm$^{-3}$ also match our constraints as long as they have $B/n^{1/2} \ge 0.5 {\rm \mu G~cm}^{3/2}$. Among the n=10 cm$^{-3}$ models, only those with a magnetic field parameter $B/n^{1/2}$ larger than 10 ${\rm \mu G~cm}^{3/2}$ satisfy our density limit.} Strong magnetic field would reduce the density compression factor because the gas is partially supported by the magnetic pressure\citep{DopitaS95}. In the following sections, we { will evaluate} two sets of shock models, those with n=1 cm$^{-3}$ and those with n=10 cm$^{-3}$. { It turns out that} shock models with { lower densities of} n=0.01 cm$^{-3}$ and n=0.1 cm$^{-3}$ would end up populating nearly the same area { in many of our diagnostic plots} as the n=1 cm$^{-3}$ models. 
}


\item{Conductive heating and Turbulent mixing models: }
The turbulent mixing models come from \cite{Slavin93}, with predictions only available for a few lines. { So we cannot do full justice to this category of models and one should consider our qualitative evaluation as tentative}. There are 6 models in total, for three different average temperatures and two different transverse velocities between the hot and the cold gas components. The lowest temperature model ($T=10^5$K) yields too low an \oiii/\oii\ ratio to produce the kind of line ratios we see. Thus, we only consider the models with higher temperatures ($T=10^{5.3}$K and $10^{5.5}$K). 
\end{enumerate}

\begin{figure}
\begin{center}
\includegraphics[width=0.48\textwidth]{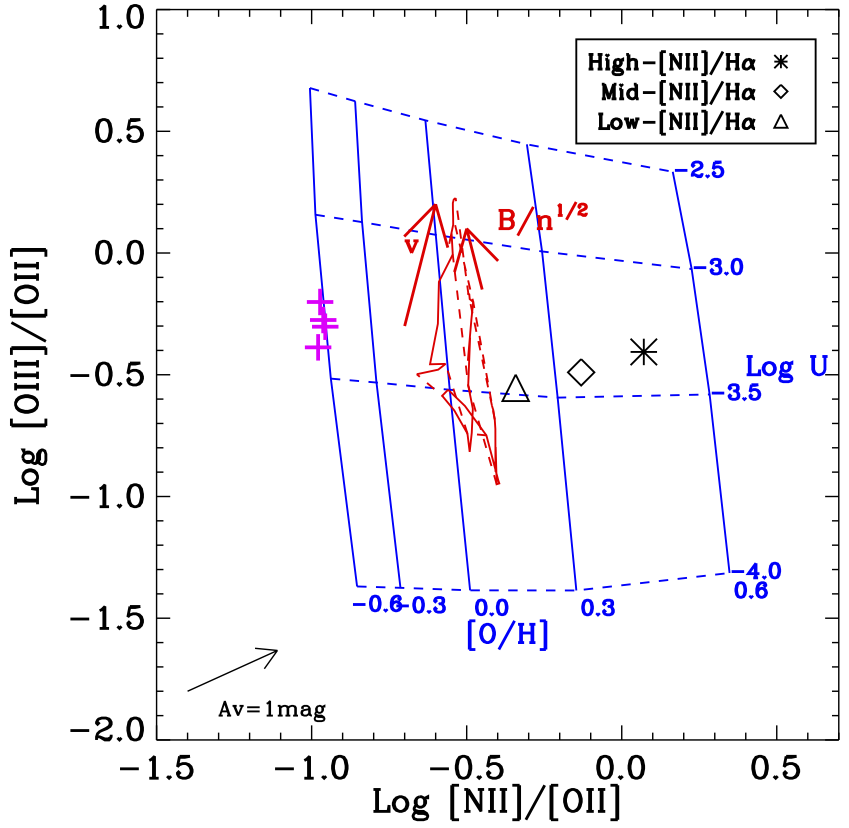}
\caption{ Comparison between the data and ionization models for the \oiiibw/\oiiw\ vs. \niibw/\oiiw\ line ratios. The blue grid at the bottom correspond to photoionization models by hot evolved stars. The solid blue lines connect models with constant [O/H] and the dashed blue lines connect models with constant ionization parameter ($\log U$). The red grid in the middle are shock models with a density of $n=10/cm^3$, in which solid red lines connect models with constant magnetic field ranging from 10 to 100${\rm \mu G~ cm}^{3/2}$ and the dashed red lines connect models with constant shock velocity, which range from 150 to 500 km/s. 
The magenta crosses indicate the turbulent mixing layer models. 
The three data points (big symbols) correspond to the three subsamples, going from the low-\nii/\hal\ (left) to the high-\nii/\hal\ (right). The error bars are not shown as they are too tiny in this plot. The data are consistent with both photoionization and shock models. { The arrow at the bottom left shows how one magnitude extinction in $A_V$ might impact the data.} Any extinction correction would therefore move the data points towards the lower left.}
\label{fig:n2o2_o3o2}
\end{center}
\end{figure}

We first { compare the data with each of the three ionization models using} the  strong-line \oiii/\oii\ vs. \nii/\oii diagram of Figure~\ref{fig:n2o2_o3o2}. In the case of the photoionization model, the data are best { reproduced} by models with $\log U\sim-3.5$ and [O/H] within $\pm0.3$ dex of the solar value. The grid shown assume a { front slab} density of n=100~cm$^{-3}$. We found that changing the density to n=10~cm$^{-3}$ hardly changes the grid at all. { These models roughly fit the observed \oi/\hal, \sii/\hal, \nii/\hal, and \oiii/\hb ratios. }
The turbulent mixing models { on the other hand} yield a rather low \nii/\oii\ ratio. In order to { reproduce the data with such a model, one would have to} significantly increase the N/O abundance ratio.
{  As for the shock models}, as mentioned earlier, { they are required to} have a sufficiently low pre-shock density or, { alternatively, a sufficiently} strong magnetic field in order to match the observed \sii\ doublet ratio. We here plot { models with} $n=10$ cm$^{-3}$, a solar metallicity and a strong magnetic field. { These succeed in matching} the \oiii/\oii\ ratio but result in a too low \nii/\oii\ ratios. { Other shock models of equal solar metallicity but} with lower pre-shock density all fall in almost the same region in this diagram, which for clarity's sake we do not show. 
To match the observed \nii/\oii\ ratios, one might invoke dust extinction or increase the { Nitrogen abundance}. 

\begin{figure}
\begin{center}
\includegraphics[width=0.48\textwidth]{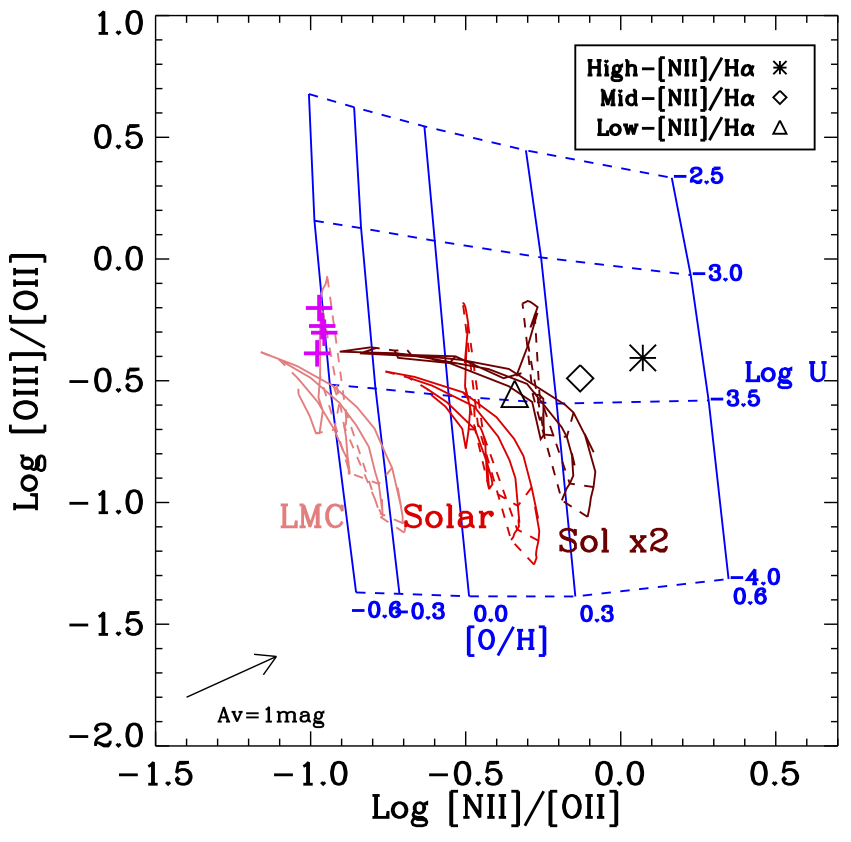}
\caption{ { This plot is similar to Figure~\ref{fig:n2o2_o3o2}, except that the three sets of shock models (with $n=1$ cm$^{-3}$) possess different metallicities: the right-most grid (dark red) corresponds to models with twice the solar metallicity, the middle grid (red) to solar metallicity, and the left-most grid (light red) to LMC metallicity.} In each grid, solid lines connect models with constant magnetic field, which range from 0.5 to 10 ${\rm \mu G~cm}^{3/2}$, with the weakest magnetic parameter yielding the largest \nii/\oii\ ratio, while dashed lines connect models with constant shock velocity, which range from 150 to 500 km/s, with the slowest shock yielding the lowest \nii/\oii\ ratio.}
\label{fig:n2o2_o3o2_metal}
\end{center}
\end{figure}

In Figure~\ref{fig:n2o2_o3o2_metal}, we show three sets of shock models with the same pre-shock density ($n=1$ cm$^{-3}$) but different sets of elemental abundances: Large Magellanic Cloud (LMC) abundances, solar abundances, and twice the solar abundances. 
{ We can clearly see the offsets in \nii/\oii\ between models}. Thus increasing the N/O abundance ratio does indeed bring the shock models { in agreement with} the data in this strong-line diagram. { Since \cite{Allen08} covered different metallicities only in the case of preshock n=1 cm$^{-3}$ models, we cannot strictly confirm how models with different pre-shock densities would behave although we expect the general trend to remain the same.}

\begin{figure}
\begin{center}
\includegraphics[width=0.48\textwidth]{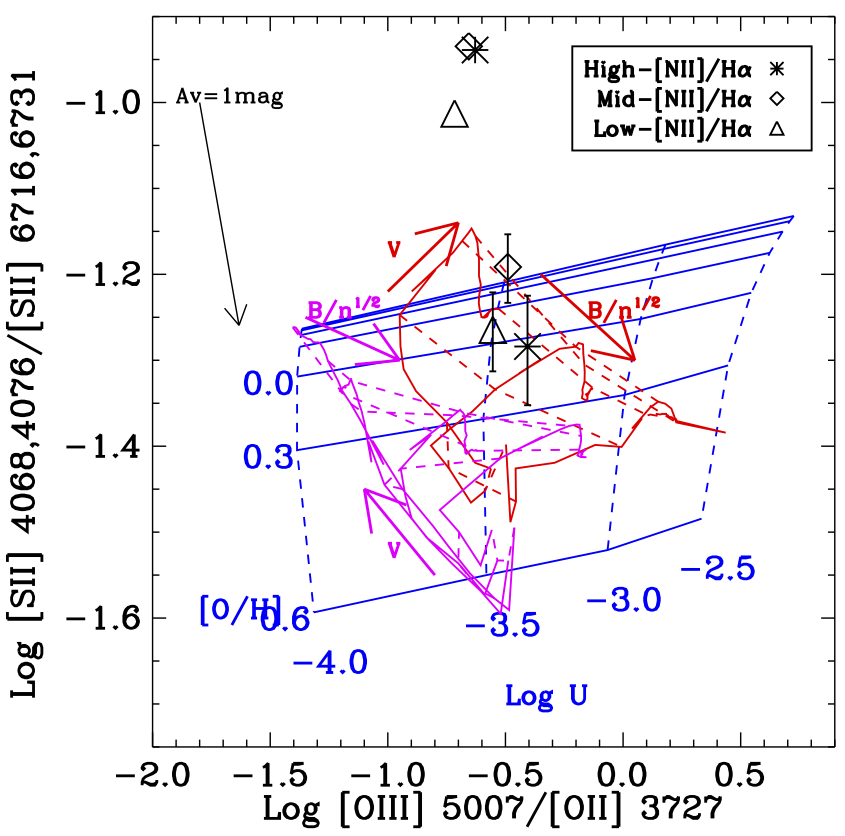}
\caption{Comparison between the data and models for the \siitw/\siiw\ vs. \oiiiw/\oiiw\ line ratios. The blue grid corresponds to photoionization by hot evolved stars with the same convention as used in Fig.~\ref{fig:n2o2_o3o2}, while the red grid is for shock models with n=10 cm$^{-3}$ and the magenta grid for shock models with n=1 cm$^{-3}$. The solid lines { connect models of the same} magnetic field parameter, with $B/n^{1/2}= 10, 30, and 100~{\rm \mu G~ cm}^{3/2}$ for the n=10 cm$^{-3}$ model and $0.5,1, 3.2, and 10 ~{\rm \mu G ~cm}^{3/2}$ for the n=1 cm$^{-3}$ model, respectively. The dashed lines represent shock velocities of $V=150, 200,300,400, 500, and 1000{\rm km/s}$. { The three large symbols with error bars represent the raw data while those without correspond to extinction-corrected ratios. The black arrow indicates the effect of 1 magnitude extinction in $A_V$.}}
\label{fig:st2_o3o2}
\end{center}
\end{figure}

The strong line diagrams are not very useful for distinguishing between photoionization and shock models. Thus, we now focus on the temperature-sensitive line ratios. We dispose of 4 temperature-sensitive line ratios. \oiitw\ and \siitw\ are detected { at a significant level} in all subsamples. We also have \niitw\ and \oiiitw\ but these lines are only detected in some subsamples. { Let's first analyze the \siitw/\siiw\ vs. \oiii/\oii\ ratios as shown in Figure~\ref{fig:st2_o3o2} and compare 
the raw data (the three data points with error bars) with both photoionization and shock models. In the case of photoionization (blue grid), the calculated \siitw/\siiw\ ratio favours subsolar metallicities. This contrasts sharply with} what we saw in the strong line plot (Figure~\ref{fig:n2o2_o3o2}). It is also surprising given our expectation of high metallicities for these massive galaxies. The n=10 cm$^{-3}$ shock models are able to match the observed ratio although shocks with lower preshock densities cannot. We note that models with a preshock density of 0.01, 0.1, and 1 cm$^{-3}$ populate approximately the same area in this plot. { For these models, decreasing the metallicity to that of LMC can raise the upper limit of the grid by 0.1 dex, which would bring it closer to some of the observed \siitw/\siiw\ ratios. If we compare the models with the reddening corrected data (the three data points without error bars), none of the models come close.}

\begin{figure}
\begin{center}
\includegraphics[width=0.48\textwidth]{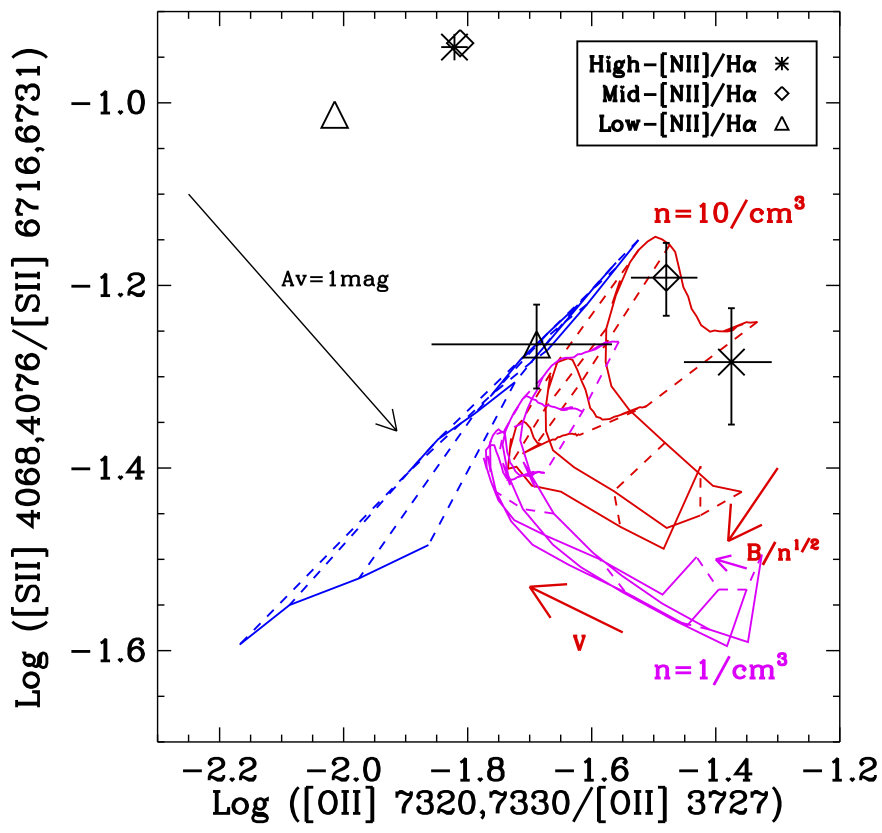}
\caption{ Comparison between the data and models for the \siitw/\siiw\ vs. \oiitw/\oiiw\ line ratios. The blue grid corresponds to photoionization by hot evolved stars with the same convention as used in Fig.~\ref{fig:n2o2_o3o2}. while the red and magenta grids correspond to the same shock models as shown in Figure~\ref{fig:st2_o3o2}. The three large symbols with error bars represent the raw data while those without correspond to extinction-corrected ratios. The black arrow indicates the effect of 1 magnitude extinction in $A_V$. The marked discrepancy between models and the dereddened ratios using the Balmer decrement suggests that the adopted extinction correction may be problematic. See text for discussion.}
\label{fig:ot2_st2}
\end{center}
\end{figure}

{ In Figure~\ref{fig:ot2_st2}, shock and photoionization models are compared with the data in a plot of the \siitw/\siiw\ vs. \oiitw/\oiiw\ line ratios. Interestingly, all models appear limited to a diagonal boundary line traced by the position of the upper leftmost photoionization models. In the case of shock models, these turn back after reaching this diagonal boundary.
This simply results from the position of the S$^+$zone, which, as compared to O$^+$, occurs deeper into the ionized cloud layer, in a region where the gas gets colder and is partially ionized, no matter whether it is photoionized or shocked. Thus, we expect in shock or photoionized models, regardless of the ionization mechanism, that the O$^+$ zone should either be co-spatial with the S$^+$ zone, or be slightly offset from it towards the warmer and more highly ionized zones. Therefore, the O$^+$ zone should either have a similar or a slightly higher temperature than the S$^+$ zone. Based on this argument, physical models are expected to populate only the diagonal domain, where the temperatures are in fact equal, or the region to the lower right, where the O$^+$ zone is hotter than S$^+$. }


In Figure~\ref{fig:ot2_st2}, we can see that the raw line ratios before extinction correction (points with error bars) appear to be consistent with both the solar metallicity, n=10/cm$^3$ shock models, and the subsolar photoionization models. The high-\nii/\hal\ subsample point can alternatively be explained by a subsolar photoionization model assuming slight extinction.  However, once extinction correction is applied, the ratios are moved too far up in the upper left of the diagram to be explained by either type of models as it would be unphysical to expect the S$^+$ zone to be hotter than the O$^+$ zone. We therefore strongly suspect that the extinction corrections derived from Balmer decrements to significantly overestimate the amount of dust reddening.

This temperature relationship between O$^+$ and S$^+$, and the opposite extinction-dependence of \oiitw/\oiiw\ and \siitw/\siiw\ make the combination of these two line ratios a useful extinction estimator. This was first pointed out by \cite{Dopita82}. Basically, in most photoionization and shock models, the two ratios should fall roughly along the diagonal zone from lower left to upper right in Figure~\ref{fig:ot2_st2}. Departures from this could be attributed to extinction. If we assume all points should fall along the diagonal line in the extinction-free case, we find the extinction in the high-, mid-, and low-\nii/\hal\ subsamples would be 0.48, 0.14, 0.0 mag in A$_{\rm V}$, respectively. Though, as shown here, very low and very high shock velocity could also make the ratios to be offset from the diagonal line. Thus, these values should be considered upper limits on dust extinction.

On the other hand, what could make the extinction estimates from Balmer decrements so unreliable? { It is worth noticing that in the case of the high- and low-\nii/\hal\ subsamples (see Figure~\ref{fig:BalmerDecrements}), smaller amounts of extinction are in better agreements with the decrements found with the higher order Balmer lines. Hence some reddening might be present coupled with a significant enhancement of \hal\ through collisional excitation. Turbulent mixing layer may play a part in overheating the partly neutral zone, increasing the Balmer decrement to as high as 5, as shown by \cite{Binette99}.
Or perhaps the extinction is non-uniform and the Hydrogen lines would tend to preferably originate from more dust-absorbed regions than the low ionization lines of O and S. Another possibility is that there may remain stellar absorption in the residual spectrum, which would make us underestimate the \hb\ strength. We do not observe, however, any trace of side wings absorption around \hb\ and so do not think this is likely either. 
This conspicuous problem with extinction is extremely interesting and need to be investigated further in the future, as we do not have a satisfying answer for the current paper.}

If we focus on shock models in Figure~\ref{fig:ot2_st2}, we find that those with n=1 cm$^{-3}$ and solar metallicity can be ruled out using these two temperature-sensitive line ratios. Even at the largest shock velocity (1000 km/s), the solar metallicity models could not produce as high a temperature in the S$^+$ zone as observed. One needs to invoke lower metallicity models to increase the temperature. However, with lower metallicity, the strong line ratio \nii/\oii\ moves further away from the observed values (see Figure~\ref{fig:n2o2_o3o2_metal}. Therefore, n=1 cm$^{-3}$ models cannot match both \siitw/\siiw\ and \nii/\oii\ simultaneously, and are ruled out.

In the case of the photoionization models, they similarly cannot match both the strong line ratios and the temperature-sensitive line ratios simultaneously using the same metallicity. Strong line ratios favour super solar metallicity while the O$^+$ and S$^+$ coronal-to-strong line ratios favour subsolar metallicity. The temperatures indicated by these line ratios might even be too high to be adequately fit using such photoionization models. So far, the only models that survive these data tests are the shock models with n=10 cm$^{-3}$ and strong magnetic parameters. We will see below that these face challenges from the other temperature-sensitive line ratios. 

\begin{figure}
\begin{center}
\includegraphics[width=0.45\textwidth]{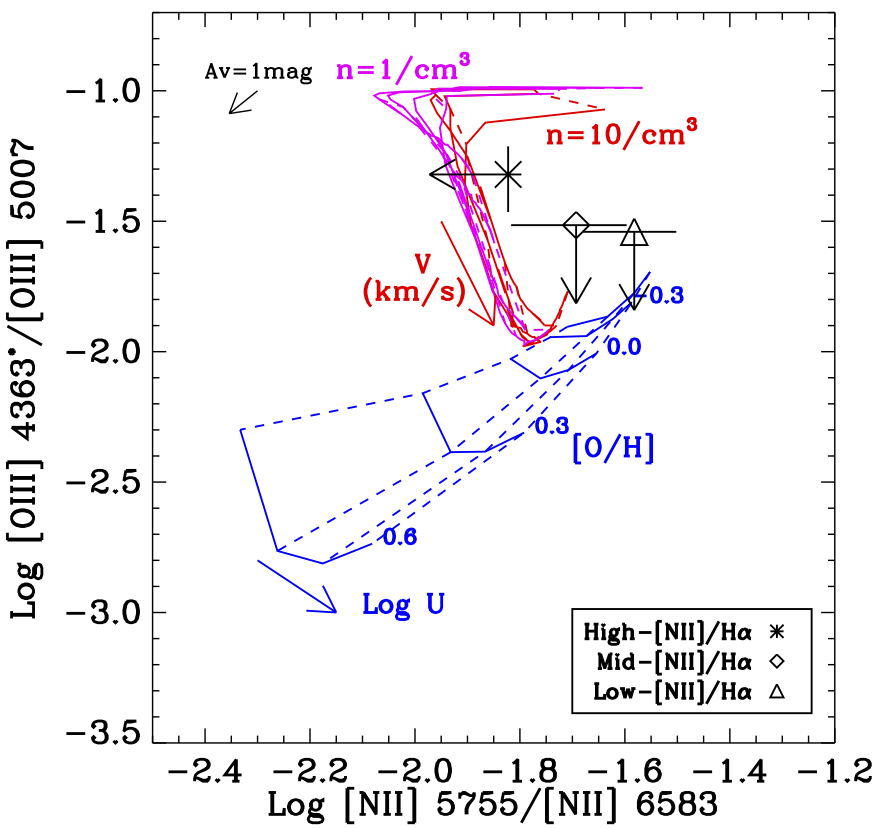}
\caption{Comparison between data with both types of ionization models in a diagram of \oiiitw/\oiiiw\ vs. \niitw/\niiw. See caption of Figure~\ref{fig:st2_o3o2} for the definition of the grid lines.  The large symbols represent our three subsamples. The arrows indicate $2\sigma$ limits when the line is undetected, otherwise, it is shown using $\pm1\sigma$ error bars. Note that for the photoionization model shown, we have added the \feiibw\ flux to the \oiiitw\ line in order to take into account possible contamination by \feii. Its contribution is significant only at the low temperature end (i.e. high metallicity and low ionization parameter). 
}
\label{fig:o3t_n2t_shock10}
\end{center}
\end{figure}

\begin{figure}
\begin{center}
\includegraphics[width=0.45\textwidth]{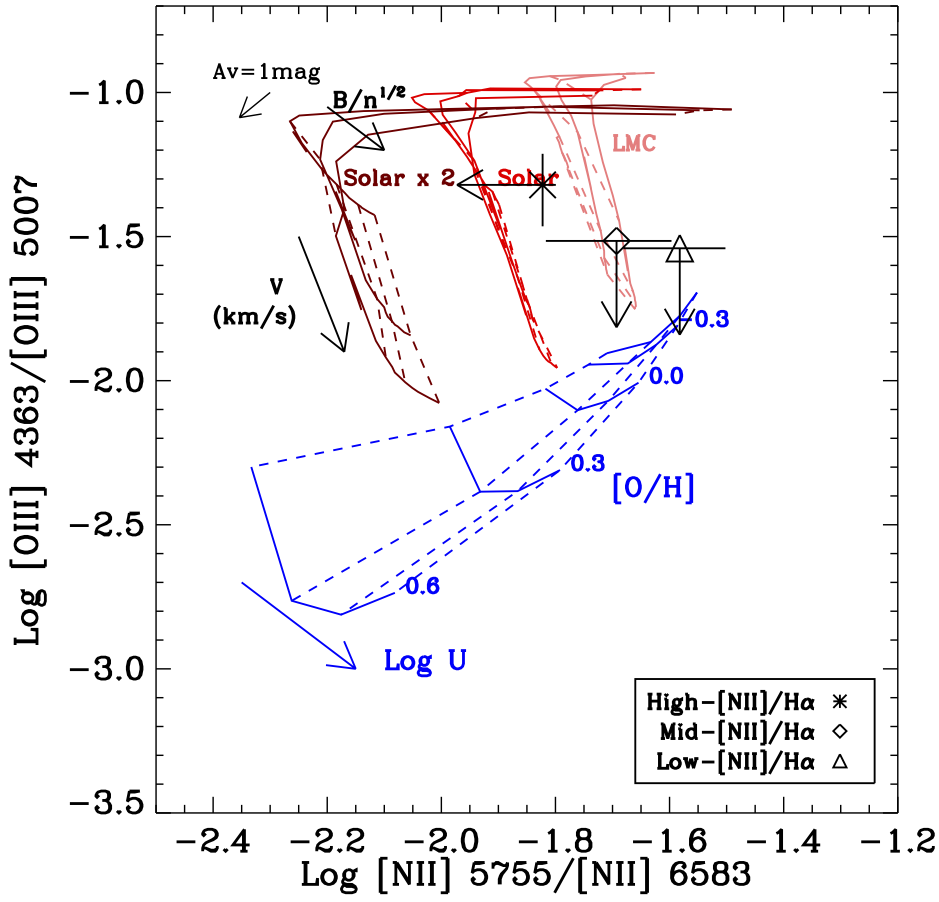}
\caption{Comparison between the data with the ionization models in a diagram of \oiiitw/\oiiiw\ vs. \niitw/\niiw. The three shock grids shown have a preshock density of n=1/cm$^3$ at three different abundance patterns. The light red grid (top right) is for the LMC abundance, the red grid (top middle) is for the solar abundance, and the dark red (top left) is for twice the solar abundance. The large symbols represent the data of the three subsamples { adding a pointing arrow when the line is undetected to indicate a $2\sigma$ limits or adding a $\pm1\sigma$ error bar when detected.} Note that for the photoionization model predictions (in blue), we have added the \feiibw\ flux to the \oiiitw\ line to take into account potential contamination by \feii. Its contribution is only significant at the low temperature end (high metallicity and low ionization parameter). 
}
\label{fig:o3t_n2t_n1_metal}
\end{center}
\end{figure}

Let us now compare the data with the models in a diagram of \oiiitw/\oiiiw\ vs. \niitw/\niiw\ line ratios, as shown in Figure~\ref{fig:o3t_n2t_shock10}. The \oiiitw\ line is only detected above 3$\sigma$ in the high-\nii/\hal\ subsample, so only for this bin is the measurement shown with $\pm1\sigma$ error bars. For the other two subsamples, we instead show the $2\sigma$ upper limits using a pointing arrow. Similarly, for the \niitw\ line, which is undetected in the high-\nii/\hal\ subsample, is here shown using a $2\sigma$ upper limit pointing arrow.
In this figure, both line ratios reveal some inadequacies in models. 
We see that the n=10 cm$^{-3}$ or the n=1 cm$^{-3}$ solar metallicity shock models, for instance, cannot match the \niitw/\niibw\ line ratios of the mid- and low-\nii/\hal\ subsamples, although they provide a good match to the marginally-detected \oiiitw\ line strength of the high-\nii/\hal\ subsample. 
If this marginal detection turned out real, the derived temperature would also come out too high to be fit by photoionization models at any metallicity considered.


All shock models discussed above are for solar metallicity. 
In Figure~\ref{fig:o3t_n2t_n1_metal}, we show how metallicity would affect the line ratios predicted by shock models. They suggest that the three \nii/\hal\ subsamples probably have different metallicities, increasing from the low to the high \nii/\hal\ bin. 
The N$^+$ zone temperature (indicated by the X-axis) turns out very sensitive to metallicity, as expected. The models show that metal-rich environment yields a lower temperature (i.e. a lower \niitw/\niibw\ ratio). The O$^{++}$ zone temperature does not appear to vary much between the three metallicities, but it is sensitive to magnetic field and shock velocity. { We find promising that a metallicity gradient in shock models could account for the observed trend across the three subsamples. However, the low-\nii/\hal\ data point shows a higher N$^+$ zone temperature than the lowest metallicity shock models (the LMC models for instance, or the SMC models which are not plotted). It would be difficult to imagine that quiescent massive galaxies could be of lower metallicity than the LMC. In addition, the indication in favour of a subsolar metallicity conflicts with that shown by  the observed \nii/\oii\ ratios which appear to favour a supersolar metallicity (Figure~\ref{fig:n2o2_o3o2_metal}). Therefore, even when taking into account the interesting possibility of metallicity variations across subsamples, there still remain discrepancies overall between the data and the shock models.}

\section{Conclusions}
Using the spectra stacking technique, we have obtained high signal-to-noise emission-line spectra of quiescent red sequence galaxies from the SDSS. After careful continuum subtraction using a control sample without emission lines, we have detected multiple temperature-sensitive coronal emission lines or provided meaningful upper limits. 

{ From reliable measurements of} \siitw, \oiitw, and \niitw\ lines, we { inferred that the emission from these ions was from regions} with a temperature of around $10^4$K. The measured \siitw/\siiw\ and \oiitw/\oiiw\ line ratios provide us with interesting constraints on the levels of extinction which are much lower than the values estimated from Balmer decrements. If the extinction values estimated from Balmer decrements {  turn out to be the correct ones}, they would lead to unphysical temperature relationship between the S$^+$ and O$^+$ emission zones. The unresolved discrepancy between the two extinction estimates is one of the puzzles we uncovered in this paper.

The \siibw/\siiaw\ ratio indicates { an electron density of} the line-emitting S$^+$ gas { that is smaller than or of the order of} 100 cm$^{-3}$. This result was used to set the physical conditions for our photoionization models { or to limit shock models either to those characterized} with a very low preshock density $n\le 1$cm$^{-3}$ and/or those with a strong magnetic field, since only such shock models can reproduce the observed \sii\ doublet ratios.

We compared the temperature-sensitive line ratios with model predictions for three ionization mechanisms: photoionization by hot evolved stars, radiative shocks, and turbulent mixing layers. We found that neither the photoionization models nor the shock models can simultaneously explain all the line ratios we observe. Photoionization models with solar and supersolar metallicities can account for the strong line ratios. { On the other hand,} all of the temperature-sensitive line ratios indicate high temperatures, { which would imply} subsolar metallicities. The marginally-detected \oiiitw\ line in the high-\nii/\hal\ sample, if it is real, would indicate too high a temperature to be fit at all by photoionization models. Shock models, which have one more degree of freedom than photoionization models, similarly cannot explain all the line ratios. Among models that { reproduce the observed} low S$^+$ density, those with solar metallicity, n=10 cm$^{-3}$, and strong magnetic field can match the observed temperatures in O$^+$, S$^+$, and O$^{++}$ zones, but fail to match the high temperature of the N$^+$ zone. Lower pre-shock density ($n\le 1 {\rm cm}^{-3}$) shock models require significantly subsolar metallicities to match the temperature of the N$^+$ and S$^{+}$ zones, but would require supersolar metallicity in order to match the \nii/\oii\ ratios.  The main discrepancy is between the observed high \nii/\oii\ ratios and the relatively hot temperatures we derived from coronal lines of singly ionized ions. Neither photoionization nor shock models could { reproduce both temperatures observed}. Given that both photoionization and shock models are failing to account for all observed line ratios for similar reasons, the combination of the two processes would face { similar difficulties.}
{ We could not do full justice to turbulent mixing models given that model predictions were only available for a few lines. }



Our work illustrates the powerful constraints provided by temperature-sensitive line ratios. They reveal significant discrepancies between data and models. Although the interpretation { of our data might be questioned on the ground of the inevitable level of uncertainty associated with} the stacking process whereby different galaxies are averaged together, it encourages us to push deeper with { deeper observations that would aim at detecting the reported lines} in individual galaxies or, { if a significantly improved S/N was achieved}, would allow us to reduce the number of galaxies required for detecting them. Our work also highlights the need of more detailed and consistent modeling that would provide us with more stringent comparisons with observations when better data become available. 
 






\section*{Acknowledgements}
I thank the referee, Luc Binette, whose comments helped to significantly improve this paper. I am also grateful for the hospitality of the Tsinghua Center for Astrophysics at Tsinghua University during an extended visit.  RY acknowledges the support of NSF Grant AST-1715898. 

    Funding for the SDSS and SDSS-II has been provided by the Alfred P. Sloan Foundation, the Participating Institutions, the National Science Foundation, the U.S. Department of Energy, the National Aeronautics and Space Administration, the Japanese Monbukagakusho, the Max Planck Society, and the Higher Education Funding Council for England. The SDSS Web Site is http://www.sdss.org/.

    The SDSS is managed by the Astrophysical Research Consortium for the Participating Institutions. The Participating Institutions are the American Museum of Natural History, Astrophysical Institute Potsdam, University of Basel, University of Cambridge, Case Western Reserve University, University of Chicago, Drexel University, Fermilab, the Institute for Advanced Study, the Japan Participation Group, Johns Hopkins University, the Joint Institute for Nuclear Astrophysics, the Kavli Institute for Particle Astrophysics and Cosmology, the Korean Scientist Group, the Chinese Academy of Sciences (LAMOST), Los Alamos National Laboratory, the Max-Planck-Institute for Astronomy (MPIA), the Max-Planck-Institute for Astrophysics (MPA), New Mexico State University, Ohio State University, University of Pittsburgh, University of Portsmouth, Princeton University, the United States Naval Observatory, and the University of Washington.


\bibliographystyle{yahapj}
\bibliography{astro_refs}

\bsp
\label{lastpage}
\end{document}